\documentclass[reprint,aps,pre]{revtex4-1}
\usepackage{graphicx}
\usepackage{dcolumn}
\usepackage{amsmath}
\usepackage{amssymb}
\usepackage{amsfonts}
\usepackage{bm}
\usepackage{color}
\usepackage{tikz}
\usepackage{atbegshi}
\usepackage{subfigure}
\usepackage{wasysym}
\usetikzlibrary{shapes,patterns,arrows}
\newcommand{\be}{\begin{equation}}
\newcommand{\ee}{\end{equation}}
\newcommand{\ba}{\begin{eqnarray}}
\newcommand{\ea}{\end{eqnarray}}

\begin{document}

\author{Gianmarco Muna\`o$^1$}
\thanks{Corresponding author, email: {\tt gmunao@unisa.it}}
\author{Antonio Pizzirusso$^1$}
\author{Antonio De Nicola$^1$}
\author{Toshihiro Kawakatsu$^2$}
\author{Florian M{\"u}ller-Plathe$^3$}
\author{Giuseppe Milano$^1$}
\affiliation{
$^{1}$Dipartimento di Chimica e Biologia, Universit\`a degli Studi di Salerno,
Via Giovanni Paolo II, 84084 Salerno, Italy. \\
$^{2}$Department of Physics, Tohoku University, Aoba, Aramaki, Aoba-ku, Sendai, Miyagi 980-8578, Japan. \\
$^{3}$Eduard-Zintl-Institut f{\"u}r Anorganische und Physikalische Chemie and Center of Smart Interfaces, Technische Universität Darmstadt, Alarich-Weiss-Str. 8, 64287 Darmstadt, Germany. \\
}

\title{Molecular Structure and Multi-Body Potential of Mean Force \\ 
in Silica-Polystyrene Nanocomposites}

\begin{abstract}
We perform a systematic application of the hybrid
particle-field molecular dynamics technique [Milano {\rm et al}, J. Chem.
Phys. 2009, {\bf 130}, 214106] to study interfacial properties and potential
of mean force (PMF) for separating nanoparticles (NPs) in a melt.
Specifically, we consider Silica NPs bare or grafted with 
Polystyrene chains, aiming to shed light on the interactions among
free and grafted chains affecting the dispersion of NPs in the nanocomposite.
The proposed hybrid models 
show good performances in catching the local structure of the
chains, and in particular their density profiles, documenting the existence of the
``wet-brush-to-dry-brush'' transition. 
By using these models, the PMF between pairs of ungrafted and grafted NPs in
Polystyrene matrix are calculated.
Moreover, we estimate the three-particle contribution to the total PMF
and its role in regulating the phase separation on the nanometer scale. In particular,
the multi-particle contribution to the PMF is able to give an explanation of the complex
experimental morphologies observed at low grafting densities.
More in general, we propose this approach and the models utilized here for a molecular
understanding of specific systems and the impact of the chemical nature of the systems
on the composite final properties. 
\end{abstract}

\maketitle

\section{Introduction}

Polymer composites containing nanosized particles 
are currently the object of a quite intensive investigation, carried out 
in the general perspective of generating new potential
technologies~\cite{Glotzer:17,Kumar:17}.
The interest in them comes from the
possibility to control the effects that polymer/filler interactions have on the
polymer chains, in order to improve the macroscopic material 
properties~\cite{Koo:06,Balazs:06,Brown:08}.
Polymer nanocomposites have been deeply investigated
by theoretical approaches~\cite{Hooper:04} and computer 
simulations~\cite{Ndoro:11,Cao:11,Meng:13,Martin:15}, beside a large variety of
experimental techniques (see, for instance, 
Refs.~\cite{Dukes:10,Xie:13,Tang:14}). As for theoretical descriptions of 
nanocomposites, most of recent studies have been focused on the density-functional theory 
(DFT)~\cite{Patel:04} 
and on the Polymer
Reference Interaction Site Model (PRISM) theory~\cite{curro1,curro2}: 
it was found that
these approaches are able to reproduce local structure and interface properties observed in
simulations~\cite{Egorov:07,Egorov:08} and experiments~\cite{Raos:08,Jancar:10,Ganesan:14}. 
However, theoretical methods 
are typically developed for generic models and not made to take into account
the chemical structure of a given compound.
On the other hand,
atomistic simulations, while providing accurate descriptions of
nanocomposites (to quote few examples, see 
Refs.~\cite{Barbier:04,Eslami:13,Denicola:15,Karatrantos:15}), 
are typically limited to small polymer chain length and short timescales.
In order to 
consider larger, properly relaxed, systems several specific coarse-grained (CG) models
have been proposed~\cite{Reith:03,Huang:10,Muller-Plathe:12,Shen:15,Shi:17}. 
A further speed up of the simulation times can be
obtained by combining the traditional molecular dynamics (MD) approaches with a field representation
of the non-bonded interactions, following the self-consistent field theory
approach~\cite{Kawakatsu:04}: according to this description, the 
mutual interactions among particles are decoupled and replaced by  
a field representation. 
In such a way it is possible to obtain a hybrid particle-field
model~\cite{Milano:09,Milano:10}, 
allowing length and time scales large enough to
successfully characterize complex systems, including polyelectrolytes~\cite{Zhu:16}, carbon 
nanotubes~\cite{Zhao:16}, polymer melts of with large molecular weight~\cite{Denicola:14} and 
nanocomposites~\cite{Denicola:16}. A more general view on the possibility to employ hybrid models
including chemical details can be found in Ref.~\cite{Milano:17}.

In the present study, we investigate the behavior of a CG model 
of a nanocomposite constituted by Silica nanoparticles (NPs) dispersed in a Polystyrene (PS) matrix 
by means of the hybrid particle-field molecular dynamics 
representation (MD-SCF)~\cite{Denicola:16}. 
The specific choice of this system is lead by the large amount of experimental studies
characterizing structural and thermodynamic properties of this 
composite~\cite{Lan:07,Akcora:09,Chevigny:10,Chevigny:11,Sunday:12}.
The main target of this study is to provide an accurate representation of the
effective interactions among NPs in the polymer melt; this is an issue of crucial importance
for the dispersion state of the NPs, that is in turn the main factor determining 
the final properties of the polymer nanocomposite.
To this end, we first validate the MD-SCF approach by comparing our results with
previous MD investigations of the same 
system~\cite{Ghanbari:13,Voyiatzis:16,Pfaller:16,Rahimi:12,Ndoro:12,Muller-Plathe:12,Ndoro:11}, 
in particular by ascertaining the accuracy of the MD-SCF scheme in reproducing
the structural properties of polymer chains.   
Once validated our approach, we use the proposed method for calculating the potential of mean
force (PMF) between a pair of NPs (bare or grafted with further PS chains) as a function of 
chain length and grafting density. Such an issue has been 
investigated in the past by means of several approaches, including DFT~\cite{Yeth:11} and PRISM 
theory~\cite{Hooper:04,Martin:13} and MD 
simulations~\cite{Cerda:03,Smith:03,Marla:06,Smith:09,Loverso:11,Meng:12,Baran:17}. 
In particular, previous numerical investigations have highlighted the 
effects due to the strength of attractive interactions among polymer and
NPs~\cite{Smith:03,Marla:06}, the role played by  
the NP size and curvature on the PMF~\cite{Loverso:11,Cerda:03}, 
the importance of the ratio between
free and grafted chain lengths~\cite{Smith:09}  
and the influence of grafting density and polymer matrix
length on the effective NP-polymer interactions~\cite{Meng:12}. 
However, all these simulation works
have been generally devoted 
to study generic models for polymers and NPs, based on pearl-necklace 
or bead-spring representations; here we deal with a specific and realistic 
model of nanocomposite, where the chemical detail is properly taken into account.
In addition,  
we have explicitly included in the calculation of the PMF the
contribution due to multi-particle effects such as three-body interactions: 
the importance of multi-particle terms
for a proper description of the PMF has been already 
highlighted in experimental works (see, for instance, 
Ref.~\cite{Akcora:09}) when facing the issue of self-assembly of
grafted NPs in polymer melts. In the same context, previous theoretical works
based on the DFT~\cite{Yeth:11} have also proved that
three-body interactions cannot be neglected in a proper description of the
PMF between NPs in polymer melts, since their contribution can be significant.
So far, 
explicit calculations of multi-body contributions to the total PMF by means of molecular simulations
have been performed mainly
for studying the formation of small clusters of nonpolar solutes in water~\cite{Rank:97},
and in particular the effects due to cooperativity in hydrophobic 
association~\cite{Czaplewski:00,Czaplewski:03}.  
An accurate treatment of the multi-body effects also in simulation studies of polymer nanocomposites
is hence highly desirable. 
Finally, molecular weights up to 1000 monomeric units
in a single PS chain have been considered, to make the system as 
realistic as possible. 

Summarizing, in the present work we:
I) validate MD-SCF predictions for the local structure of PS chains
by comparing our results with previous MD studies;
II) exploit our approach for studying the PMF between two NPs in a PS melt as a function
of chain length and grafting densities;
III) estimate the three-body contribution to the total PMF,
discussing this effect in comparison with simulation and experimental data on Silica NPs
dispersed in PS and gaining more detail inside the effective interactions between NPs.

All details concerning the CG models and the simulation procedure are provided in the {\bf Model systems} 
and {\bf Methods} sections respectively,
while results concerning the above said points are presented in the 
{\bf Results and Discussion} section. Final remarks are given in the {\bf Conclusions} section.
\begin{figure}[t!]
\begin{center}
\includegraphics[width=9.0cm,angle=0]{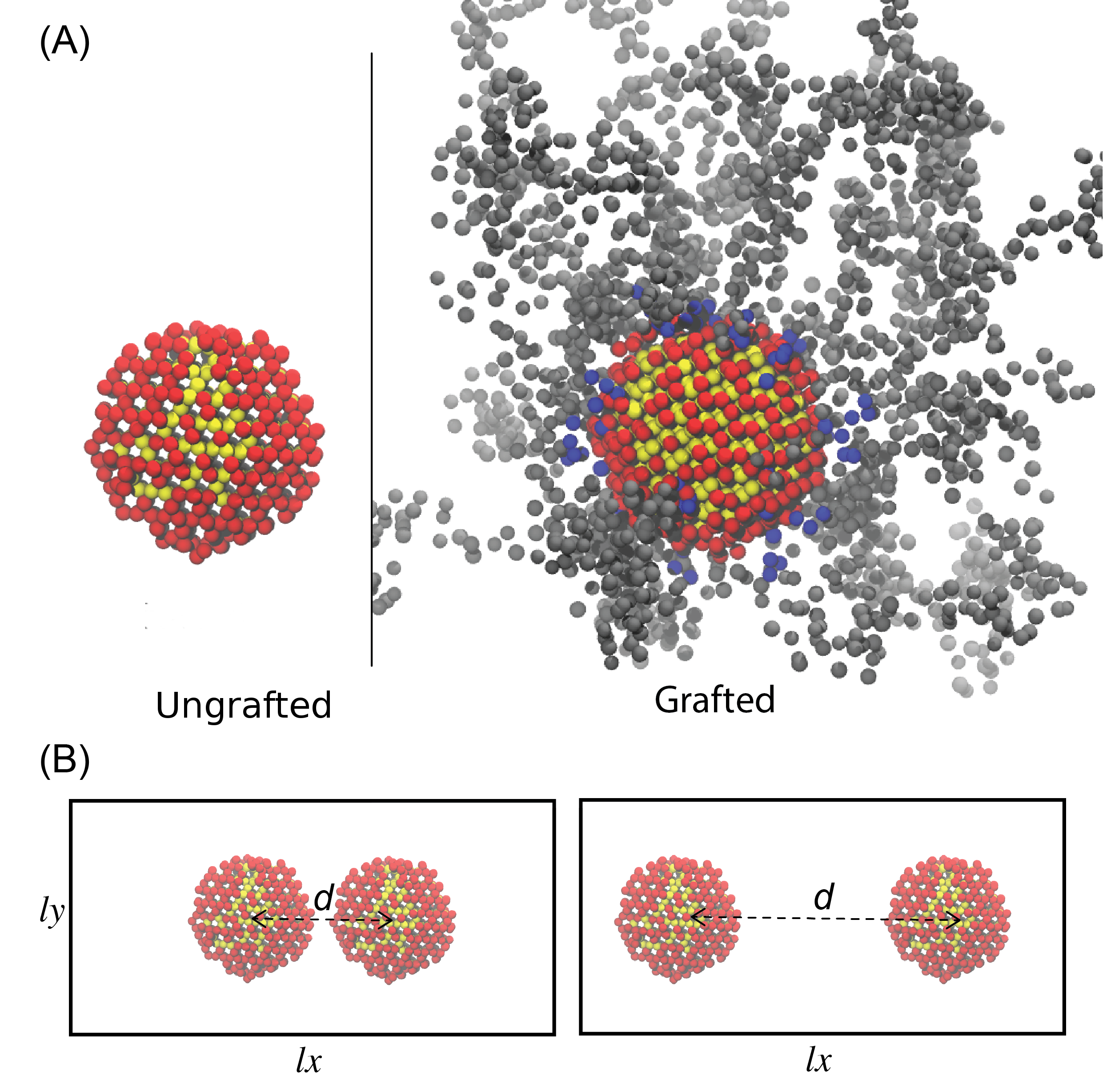}
\caption{Panel (a):
cartoons of typical ungrafted and grafted NPs
investigated in this work; NP beads are depicted in red and yellow
colors, linkers in blue and polymer chains in grey. Panel (b): sketch of the
procedure for calculating the two-body PMF upon increasing the distance between
the NPs.}
\label{fig:models}
\end{center}
\end{figure}

\section{Model systems}
The CG models that we adopt for studying Silica
NPs embedded in PS melts have been introduced by Qian and
coworkers~\cite{Qian:08} and successfully implemented in
Refs.~\cite{Muller-Plathe:12,Denicola:16}:
each repeating units of atactic PS is represented
as one bead located at the repeating unit center-of-mass. Two different
beads (R and S) account for the chirality of the asymmetric carbon.
As for the NP,
the effective bead is centered on the silicon atom position
and represents one ${\rm SiO_2}$ unit. The total mass of the hydrogen atoms
is distributed over all the CG beads of the
NP. All simulated NPs have a diameter of 4 nm and contain 873 beads.
In addition, we also consider systems where some PS chains are grafted to
the NP surface. Following the prescription by Ghanbari and
coworkers~\cite{Muller-Plathe:12}, each grafted chain is attached to the NP
surface through a linker unit, that is divided into four CG beads of two kinds
with the same mass. A representative snapshot of ungrafted and grafted
NPs is shown in
Fig.~\ref{fig:models} (panel a), along with the schematic procedure for
calculating the two-body PMF (panel b).
Further details on this CG representation and
its validation can be found in
Refs.~\cite{Muller-Plathe:12,Denicola:16}.

It is worth noting that, as already pointed out in Ref.~\cite{Denicola:16},
the periodic interstitial structure
of the NP surface gives rise to
an oscillating
density field inside the NP core; as a consequence, the incompressibility
condition can lead to an unphysical presence of PS chains inside the NP.
This is true in particular for ungrafted NPs, whereas for grafted NPs
the presence of the external chains precludes the free chains to get close to
the NP surface. To counteract this effect,
the NP density is described by an analytical effective
field: such an approach
has been applied also in SCF theory to deal with excluded volume interactions
of solid NPs~\cite{Sides:06}. Here
we model the NP density field as a combination of two different splines:
a first,
quadratic function describes the density field inside the NP core, and a second,
cubic function prevents the external chains from overlapping the NP.
These functions can be calculated by fixing four parameters:
the NP radius $r_0$, the maximum value of the density field inside the NP 
($\phi_{core}$), the value of the density field at the NP surface ($\phi_{surf}$) and  the interval
$\delta r$ giving the width of the NP density profile ({\it i.e.} how fast the NP
density field goes from $\phi_{surf}$ at $r = r_0$ to zero at
$r = r_0 + \delta r$). In our model, $r_0=2$, $\phi_{core}=100$,
$\phi_{surf}=2$ and $\delta r=0.50$.
Spline coefficients are obtained by
imposing these conditions and continuity of NP density and
its derivative.
Further details can be found in Ref.~\cite{Denicola:16}.

\section{Methods}

The simulation approach that we adopt in the present study for studying CG systems 
is based on a combination of a standard molecular
dynamics approach and a self-consistent field theory ~\cite{Kawakatsu:04} 
for the calculation of non-bonded potentials. The resulting scheme is known as
hybrid particle-field model~\cite{Milano:09}: according to such an approach,
the hamiltonian of a system comprised by $M$ molecules can be split as:
\begin{equation}\label{eq:H}
\hat{H}(\Gamma)=\hat{H_0}(\Gamma)+\hat{W}(\Gamma)
\end{equation}
where $\Gamma$ represents a point in the phase space and the symbol \hspace{0.01cm} $\hat{}$
\hspace{0.01cm} indicates that a given quantity is a function of the microscopic state
corresponding to $\Gamma$. In Eq.~\ref{eq:H}, $\hat{H_0}(\Gamma)$ is the
Hamiltonian of a system with molecules experiencing only intramolecular
interactions, whereas $\hat{W}(\Gamma)$ is the contribution due to the other
non-bonded interactions.
The latter can be calculated as an external potential $V({\bf r})$ on single 
particles, which is due to the density field. The details of the derivation 
of $V({\bf r})$ can be found elsewhere~\cite{Milano:09}. According to this 
derivation, the mean field solution for the
potential acting on a particle of type $K$ at position $r$, $V({\bf r})$ is: 
\begin{equation}\label{eq:pot}
V_K({\bf r})=k_B T \sum_{K'} \chi_{KK'} \Phi_{K'}({\bf r}) +
\frac{1}{\kappa}(\sum_K \Phi_K ({\bf r}) -1)
\end{equation}
where $k_B$ is the Boltzmann constant, $T$ is the temperature, 
$\chi_{KK'}$ are the mean field parameters for the interaction of a 
particle of type $K$ with the density field due to particles of type $K'$ and 
the second term on the right-hand
side of Eq.~\ref{eq:pot} is the incompressibility condition,
$\kappa$ being the compressibility. Also, $\Phi_K ({\bf r})$ and $\Phi_{K'}({\bf r})$
are the density functions of the CG beads of type $K$ and $K'$, respectively.
The value of $\chi_{KK'}$ has been fixed to 5.25, in agreement with a 
previous MD-SCF study of the same model~\cite{Denicola:16}.

All CG simulations have been performed by means of the OCCAM code, whose
details can be found in Ref.~\cite{Occam}; in particular,
MD-SCF simulations have been performed in the NVT ensemble, with the
temperature (fixed at 590 K) controlled by the Andersen thermostat and a
time step of 4 fs. All particles have been 
enclosed in a cubic simulation box with
periodic boundary conditions: values of box sizes are reported in Tab.~\ref{tab:scf}.
In the implementation employed here, we have 
divided
the simulation box into cubic cells of size  $l=0.6$ and 1.2 nm. 
The total force acting on each particle is the sum of intramolecular
interactions (obtained by means of tabulated potentials) 
and contributions due to its interactions with the density fields. 

\begin{table*}
\begin{center}
\caption{Nanocomposite systems considered in MD-SCF simulations. The grafting density
is given in chains/nm$^2$ and the box length in nanometers. There is one NP for each
system.}
\label{tab:scf}
\begin{tabular*}{1.0\textwidth}{@{\extracolsep{\fill}}cccccccccccc}
\hline
\hline
& \hspace{-0.5cm} Grafting density &  Grafted chains &
Grafted chains &  Free chains & Free chains & Box length \\
& \hspace{-0.5cm} ($\rho_g$) & ($N_g$) & length ($L_g$) & ($N_f$) & length ($L_f$) & ($L_b$) \\
\hline
& \hspace{-0.5cm} 0.5  &  25 &  80 &  808 &  20 & 15.20 \\
& \hspace{-0.5cm} 0.5  &  25 &  80 &  24 &  2400 & 22.40\\
& \hspace{-0.5cm} 1.0  &  50 &  20 &  177 &  20 & 9.70 \\
& \hspace{-0.5cm} 1.0  &  50 &  80 &  708 &  20 & 15.18 \\
& \hspace{-0.5cm} 1.0  &  50 &  80 &  177 &  80 & 15.07 \\
& \hspace{-0.5cm} 1.0  &  50 &  80 &  88 &  160 & 15.04 \\
& \hspace{-0.5cm} 1.0  &  50 &  80 &  24 &  2400 & 22.55 \\
\hline
\end{tabular*}
\end{center}
\end{table*}
\begin{figure*}[t!]
\begin{center}
\includegraphics[width=14.0cm,angle=0]{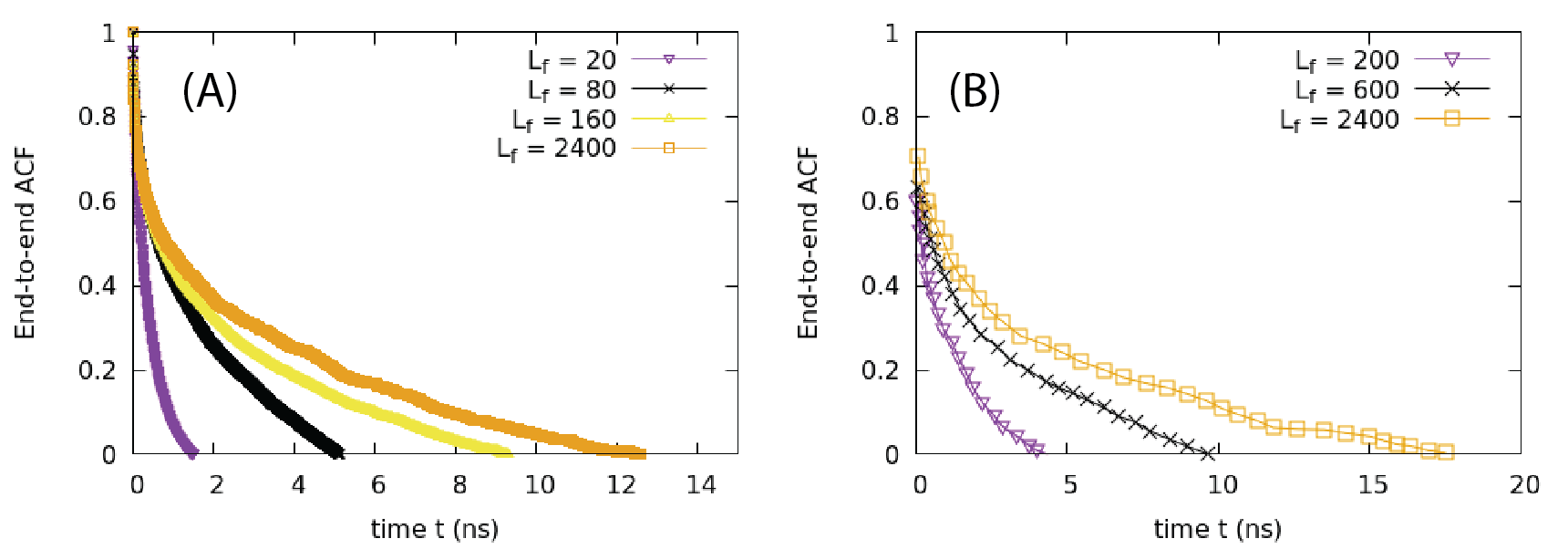} \\
\hspace{0.7cm}
\includegraphics[width=7.0cm,angle=0]{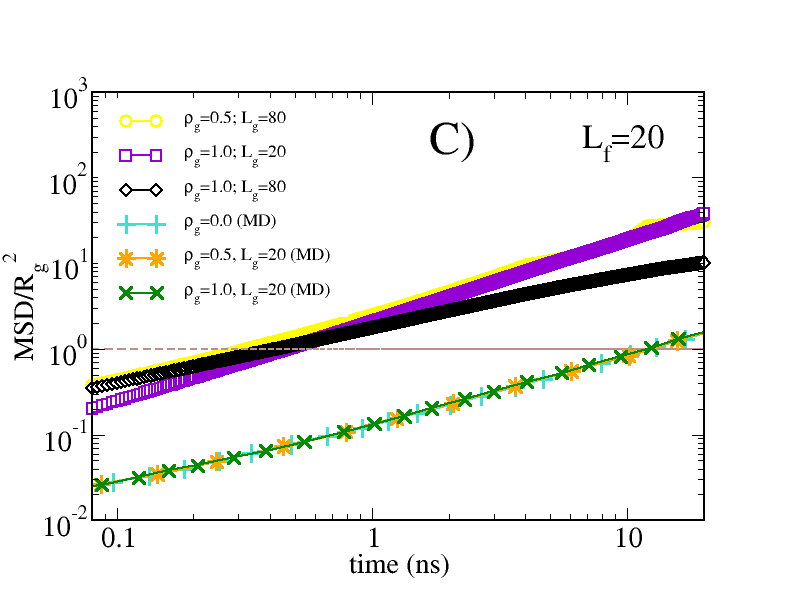}
\includegraphics[width=7.0cm,angle=0]{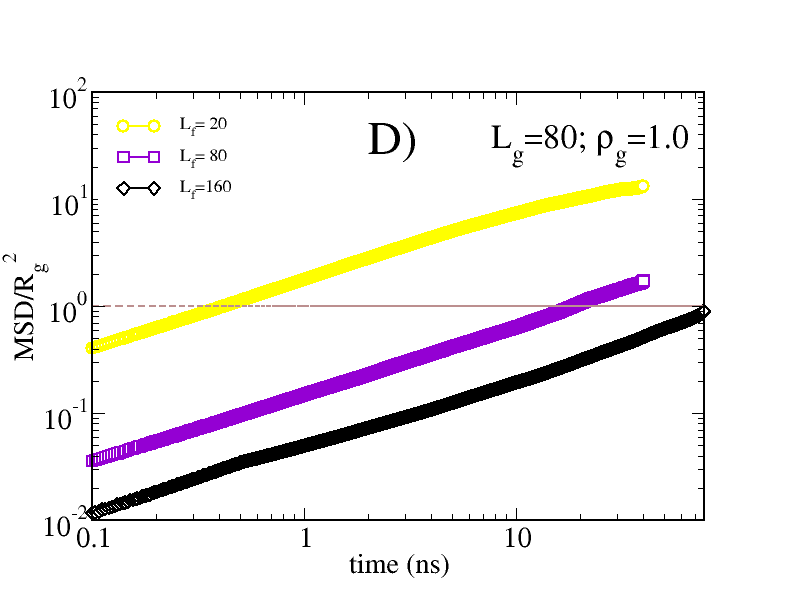}
\caption{Panel (a): autocorrelation function of the end-to-end vector for free PS chains
embedding a grafted Silica NP with $\rho_g = 1.0$ chains/nm$^2$ and $L_g = 80$.
Panel (b): same for PS chains embedding
a ungrafted Silica NP from Ref.~\cite{Denicola:16}. Values of
$L_f$ are in the legends.
Panels (c), (d): mean square displacement (MSD) of the
center of mass of free chains in units of the gyration radius ($R_g$) as
a function of the time in a logarithmic scale. In panel (c)
we report results for free chains of $L_f=20$
and various $\rho_g$ and $L_g$ obtained by MD-SCF approach and compared with
atomistic MD simulations from Ref.~\cite{Ndoro:12}.
In panel (d) we show results for
$\rho_g = 1.0$ chains/nm$^2$, $L_g = 80$ and various $L_f$ obtained by MD-SCF approach.
}
\label{fig:end}
\end{center}
\end{figure*}
As for the calculation of the two-body PMF,
for each system we have prepared
a set of 30 independent initial
configurations,
each one corresponding to a pair of NPs
placed at a fixed distance from each
other and embedded in the polymer matrix. Such initial configurations
have been built by using the Packmol program~\cite{Packmol}.
One simulation has been performed for each
configuration, where the NPs were allowed to freely rotate but
not to translate, in order to keep their mutual distance fixed.
Forces on the centers of mass of the two NPs
have been computed each 0.4 ps and
averaged over 80 ns.
Convergence has been ensured by verifying that the average values of the
forces do not change anymore up to the first significant figure.
If not explicitly reported in the figures, error bars
corresponding to standard deviations are smaller than symbol sizes
of the corresponding curves.
The resulting PMF has been calculated according to the equation:
\begin{equation}\label{eq:PMF}
U(r)=-\int_{r_{min}}^{r_{max}} \bar{F}(r) dr
\end{equation}
where $r$ is the interparticle distance, ranging in the $[r_{min},r_{max}]$
interval. In our simulations, $r_{min}=4$ nm and $r_{max}=10$ nm;
therefore $r_{min}$ is coincident with the NP diameter, while $r_{max}$
indicates an interparticle distance where the potential can be confidently
assumed equal to zero. In all simulations, distances are sampled with a step
of 0.2 nm and the numerical integration is performed by employing the trapezoidal rule.

\section{Results and Discussion}

\subsection{Relaxation times and molecular structure of polymer chains}

The main advantage of the hybrid particle-field scheme
is that non-bonded interactions
are calculated among single particles and an external potential. Since such interactions
constitute the computationally most intensive part of MD simulations,
even nanocomposites containing NPs and
polymers with high molecular weights can be well relaxed~\cite{Denicola:14,Denicola:16}.

\begin{figure*}[t!]
\begin{center}
\includegraphics[width=10.0cm,angle=0]{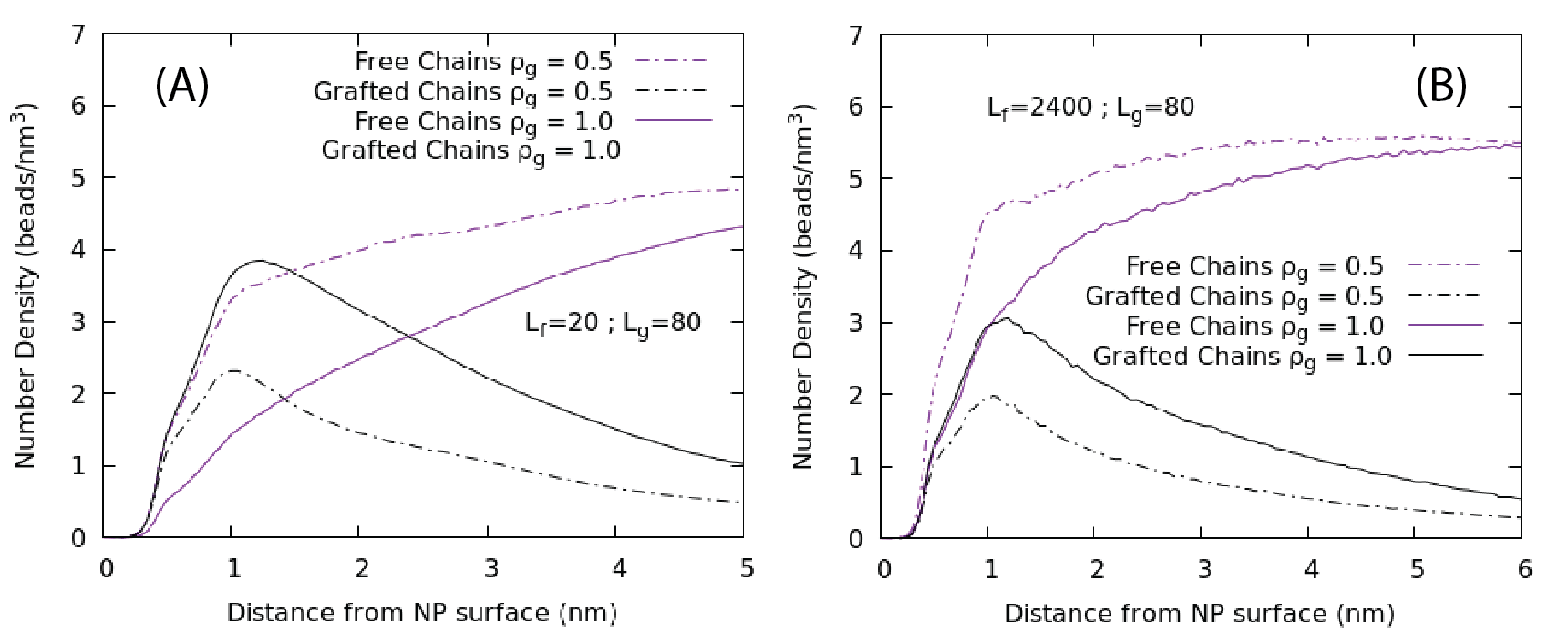}
\includegraphics[width=5.0cm,angle=0]{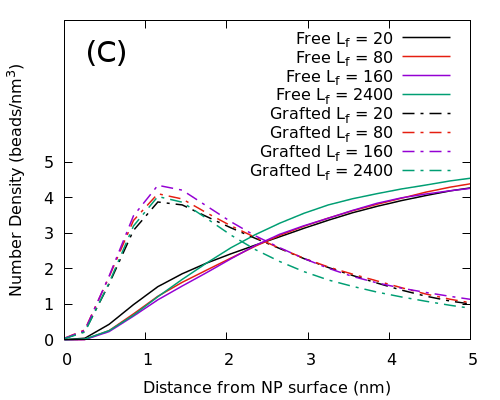}
\caption{Panels (a), (b): bead number density of free and grafted
chains for $\rho_g=0.5$ chains/nm$^2$ (dashed lines) and $\rho_g=1.0$ chains/nm$^2$ (full lines)
as a function of the distance from the NP surface,
with fixed $L_g = 80$ and two different $L_f$, namely 20 (a) and 2400
(b). The effect of increasing $L_f$ on the bead number density of the chains (for fixed $\rho_g=1.0$ 
chains/nm$^2$ and $L_g=80$) is reported in panel (c).}
\label{fig:dens}
\end{center}
\end{figure*}
The complete collection of systems investigated by means of MD-SCF simulations
is reported in Tab.~\ref{tab:scf}: in all cases there is one NP in the PS melt, grafted with other PS chains of
variable length.
We consider several values of free
and grafted chain length ($L_f$ and $L_g$, respectively) for two
different grafting densities $\rho_g$, corresponding to two different numbers
of grafted chains $N_g$. The number of free chains $N_f$ is also allowed to
vary.
Relaxation times and local structure of free and grafted chains are collectively
reported in Fig.~\ref{fig:end}: specifically, we show in the top panels the autocorrelation
function (ACF) of the end-to-end vector of
free chains of variable length embedding a single grafted NP
with $\rho_g = 1.0$ chains/nm$^2$ and $L_g = 80$. The simulation time required to obtain an
independent chain configuration is equivalent
to the time that the corresponding
end-to-end vector ACF needs to decay to zero. Even for long PS chains
($L_f=2400$) this time is less than 13 ns, which indicates the
efficiency of the MD-SCF approach. 
In order to investigate the role played by the presence of the grafted chains on the
relaxation times of the free chains we have compared our results (panel a) with previous calculations made for 
ungrafted NPs~\cite{Denicola:16} (panel b). The comparison is purely indicative, since the specific
values of $L_f$ are not the same, but it seems to suggest that the presence of grafted chains
increases the relaxation times of the free chains. A noticeable exception is found for $L_f=2400$, where this
trend is reversed: 
this can be explained in terms of
repulsive interactions among free and grafted chains, this repulsion becoming
stronger upon increasing
$\rho_g$ and $L_f$, hence giving rise to a faster chain relaxation.
\begin{table*}[t!]
\begin{center}
\caption{Radii of gyration ($R_{gf}$ and  $R_{gg}$) and end-to-end distances
(both in nanometers) of, respectively, free and grafted
chains for all systems investigated.}\label{tab:gyr}
\begin{tabular*}{1.02\textwidth}{@{\extracolsep{\fill}}cccccccccccccccc}
\hline
\hline
& $\rho_g$ &  $L_g$ &  $N_f$ &  $L_f$ &
$R_{gf}$ &  $R_{gg}$  &  End-to-end (free)
&  End-to-end (grafted)\\
\hline
& 0.5 &  80 &  808 &  20 &  0.909 $\pm$ 0.002 &
2.4 $\pm$ 0.16 &  2.039 $\pm$ 0.008 &  6.5 $\pm$ 0.5 \\
& 0.5 &  80 &   24 &  2400 & 8.06 $\pm$ 0.02 &
2.808 $\pm$ 0.004 &  10.2 $\pm$ 0.21 &  8.2 $\pm$ 0.16 \\
& 1.0 &  20 &  177 &  20 &  0.908 $\pm$ 0.005 &
0.93 $\pm$ 0.01 & 2.02 $\pm$ 0.02 & 2.25 $\pm$ 0.06 & \\
& 1.0 &  80 &  708 &  20 &  0.916 $\pm$ 0.002 &
2.41 $\pm$ 0.07 &  2.055 $\pm$ 0.009 &  6.3 $\pm$ 0.27 \\
& 1.0 &  80 &  177 &  80 &  1.99 $\pm$ 0.02 &
2.49 $\pm$ 0.03 &  4.56 $\pm$ 0.07 &  6.6 $\pm$ 0.14 \\
& 1.0 &  80 &   88 &  160 &  2.83 $\pm$ 0.05 &
2.3 $\pm$ 0.13 &  5.9 $\pm$ 0.20 &  6.2 $\pm$ 0.39\\
& 1.0 &  80 &   24 &  2400 &  8.42  $\pm$ 0.02
&  2.51 $\pm$ 0.05 &  10.5 $\pm$ 0.31 &  7 $\pm$ 0.17\\
\hline
\end{tabular*}
\end{center}
\end{table*}
\begin{figure*}[t!]
\begin{center}
\begin{tabular}{ccc}
\includegraphics[width=6.2cm,angle=0]{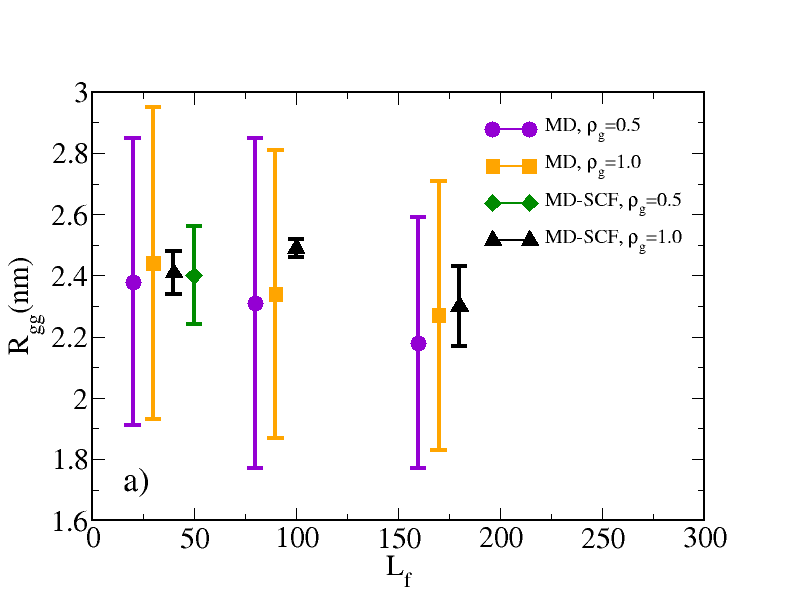} 
\includegraphics[width=6.2cm,angle=0]{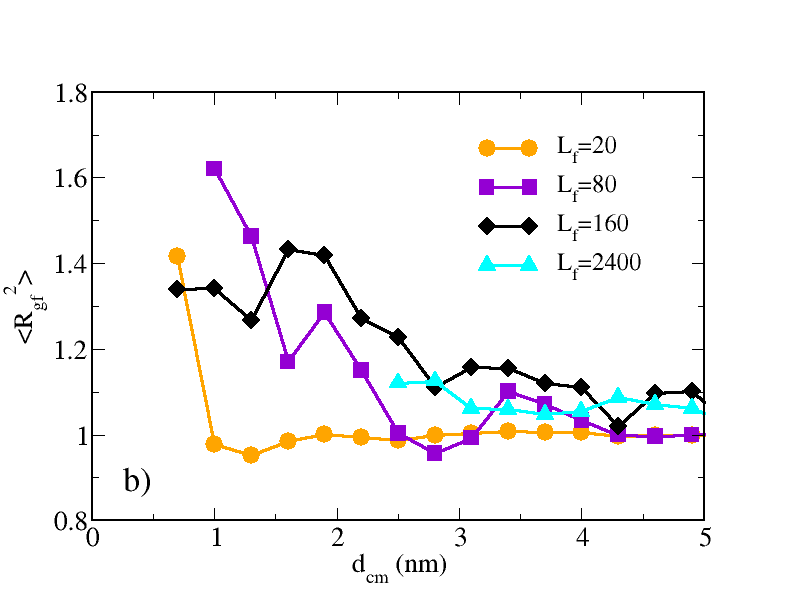} 
\includegraphics[width=6.2cm,angle=0]{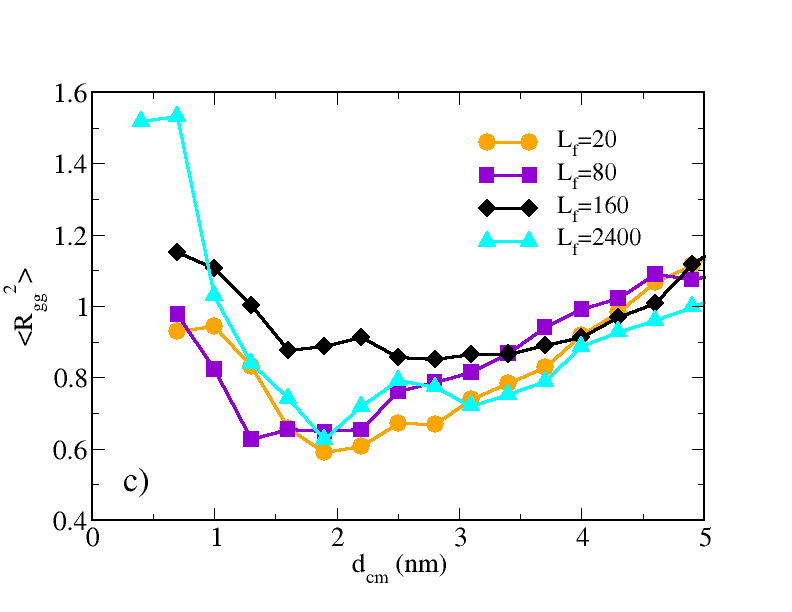}
\end{tabular}
\caption{Panel (a): Comparison among MD and MD-SCF results for $R_{gg}$
with fixed $L_g=80$ and various $L_f$. MD values are taken
from Ref.~\cite{Muller-Plathe:12} For the sake of clarity, squares, triangles and diamonds have been
shifted on the x-axis of 10, 20 and 30 units, respectively. Panel (b): squared radius of gyration
of free chains ($R_{gf}$) as a function of the distance $d_{cm}$ between
their centers of mass and the surface of the NP. Panel (c): same for
grafted chains ($R_{gg}$) as a function of the distance $d_{cm}$ between
their centers of mass and the surface of the NP. In both panels
$\rho_g=1.0$ chains/nm$^2$, $L_g=80$ and $L_f$ takes the values reported in the legends.}
\label{fig:distrib}
\end{center}
\end{figure*}
Further indication on the relaxation time of PS chains can be gained by
calculating the MSD, since, following Ref.~\cite{Denicola:16}, one can assume
that the equilibration procedure is effective if each chain moves its center
of mass by at least its gyration radius.
In the bottom panels of Fig.~\ref{fig:end}
we show the time dependence of the MSD (divided by $R_g^2$) for
a fixed $L_f=20$ with various $L_g$ and $\rho_g$
(panel c) and for fixed $\rho_g=1.0$ chains/nm$^2$ and $L_g=80$ with various $L_f$ (panel d).
In panel (c) we also report a
comparison with bulk MSD values obtained in previous atomistic MD simulations
of the same system~\cite{Ndoro:12}
for $L_f=L_g=20$ and various $\rho_g$.
For $L_f=20$ the MD-SCF approach is faster than standard MD simulations of
almost two order of magnitude in relaxing the chains: this is due both to the speed up
given by the field representation and to the adoption of CG models instead of atomistic
representations. In addition, it is worth noting that in MD-SCF simulations there is a
(slight) dependence of the MSD on the grafting density. In MD simulations
this dependence is not observed, but this is due to the fact that in the figure only the
bulk values of free chains are reported; as stated in Ref.~\cite{Ndoro:12}, the MSD
calculated for chains close to the NP surface depends on $\rho_g$, as observed also in
our results.
Finally, the increase of the MSD with $L_f$ is reported in Fig.~\ref{fig:end}d:
as already known for Gaussian polymer chains, MSD scales as $L_f^{-2}$ even in
presence of a grafted NP; according to MD-SCF approach, chains containing up to 160
beads can be relaxed within 80 ns.
Further evidence of the speed-up of hybrid simulations in comparison with the standard MD approach 
in properly relaxing polymer melts and nanocomposites can be found in Refs.~\cite{Denicola:14,Denicola:16}

The bead number density of free and grafted chains
as a function of the distance from the NP surface is reported in 
Fig.~\ref{fig:dens}. 
In particular, the effect of increasing $\rho_g$ on the density profile of
short ($L_f=20$) and long ($L_f=2400$) free chains is investigated in panels (a) and (b), 
respectively, while the role played by the free chain length is analyzed in panel (c).
It emerges that upon increasing $\rho_g$ the peak of the grafted chains distribution increases,
while that of free chains decreases; also, for low distances ($< 2.5$ nm) from the NP surface,
there is a slight drop of the bead number
density of free chains when $L_f$ increases, whereas this behavior is reversed for
grafted chains. At higher distances, the number density keeps almost constant. 
The picture is compatible with the known ``wet-brush-to-dry-brush'' transition,
observed in both experimental~\cite{Chevigny:10} and numerical~\cite{Ndoro:11,Muller-Plathe:12} 
studies of similar systems, and corresponding to the progressive expulsion of the free chains
from the grafted corona when the grafting density increases.
It is also worth noting that both the existence of a crossover distance (2.5 nm in our model)
and the dependence of the transition on $\rho_g$ and not on $L_f$ have been documented in
Refs.~\cite{Ndoro:11,Muller-Plathe:12}. 

\subsection{Chains extension and orientation}
The investigation of the chain structure is completed by analyzing their extension
and spatial arrangement.
Gyration radii and end-to-end distances of free and grafted chains for progressively higher
$\rho_g$ and $L_f$ are reported
in Tab.~\ref{tab:gyr}, where $R_{gf}$ and $R_{gg}$ label the gyration radius of
free and grafted chains, respectively.
Neither the gyration radii nor the
end-to-end distances depend on the grafting density,
since their values are practically unchanged upon
increasing $\rho_g$ from 0.5 to 1.0 chains/nm$^2$; on the other hand, $R_{gf}$ and the end-to-end
distance of free chains significantly increase upon increasing $L_f$, as can be expected.
This is not the case of $R_{gg}$ and the end-to-end distance of grafted chains,
whose values are
almost constant regardless of the specific values of $\rho_g$ and $L_f$.

\begin{figure*}[t!]
\begin{center}
\includegraphics[width=3.5cm,angle=0]{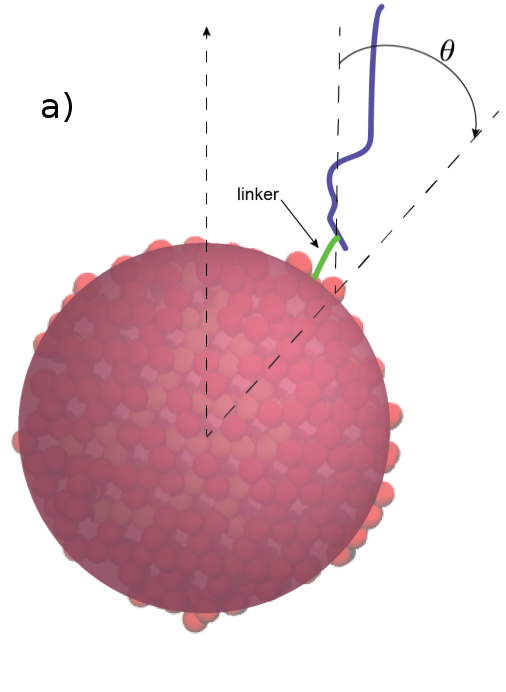}
\includegraphics[width=6.5cm,angle=0]{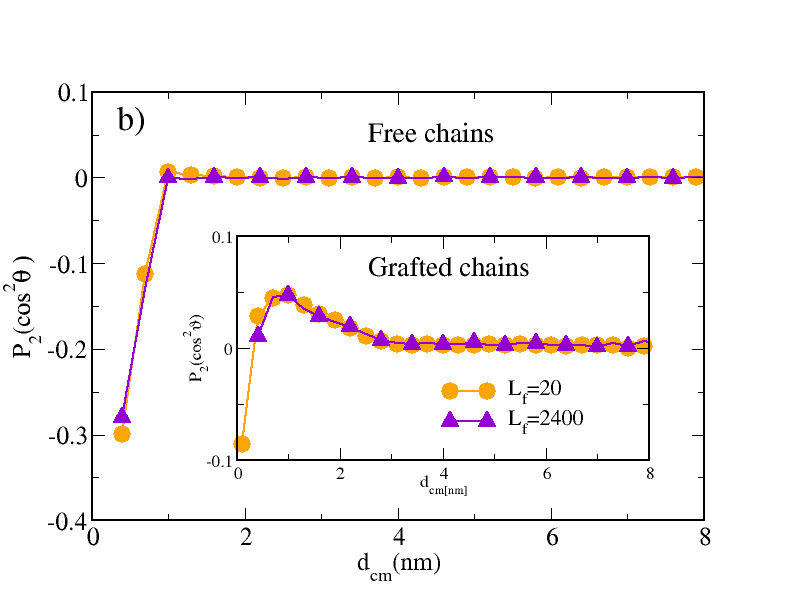}
\includegraphics[width=6.5cm,angle=0]{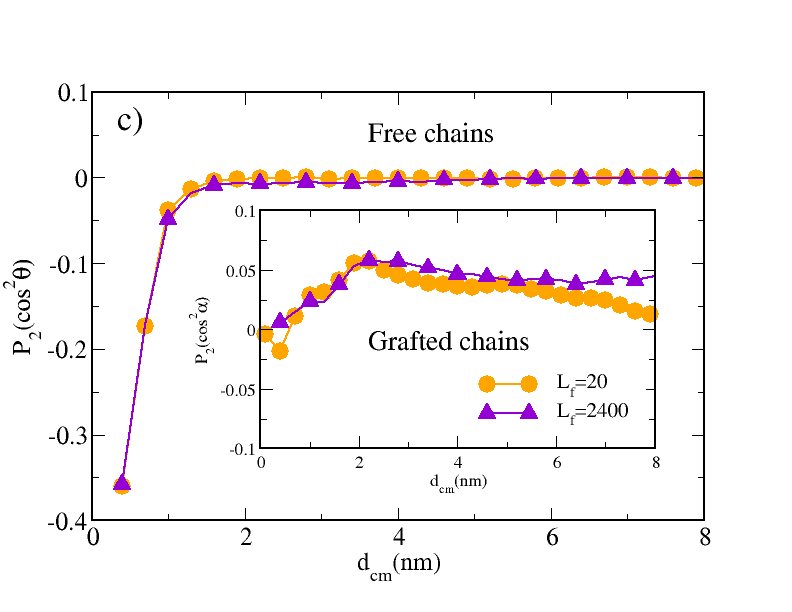}
\caption{ 
Panel (a): visual representation of the orientational angle $\theta$ between the NP-surface
normal and the vector joining two neighbouring beads of a grafted chain.
The same definition holds for free chains. Panels (b), (c): orientation angle distribution for nearest 
and next-nearest 
segment vectors of free and grafted chains (in the inset), respectively. We report
the second Legendre polynomial of $\cos(\theta)$, where $\theta$ is the angle
between the inter-segment vector and the normal of the NP surface.
In all panels
$\rho_g=1.0$ chains/nm$^2$, $L_g=80$ and $L_f$ takes the values reported in the legends.}
\label{fig:orient}
\end{center}
\end{figure*}
MD-SCF results for the radii of gyration of grafted chains are assessed against
MD simulations of the same CG model~\cite{Muller-Plathe:12} in Fig.~\ref{fig:distrib}a:
it turns out that
MD-SCF results are in agreement with MD data in
providing the  independence of $R_{gg}$ on both $\rho_g$ and $L_f$. In addition,
it is worth noting that the error bars of MD data
are systematically higher than those of MD-SCF simulations, this suggesting the capability
of this approach to provide more precise estimates of structural parameters.

The dependence of the gyration radius on the distance $d_{cm}$ between the centers of mass
of the chains and the surface of the NP
is shown in panels (b) and (c) of Fig.~\ref{fig:distrib}. $R_{gf}$ (panel b)
remarkably increases for $L_f=20$ at short $d_{cm}$:
this circumstance suggests that in such conditions the free chains are flattened against
the NP surface. This effect is observed only for small $L_f$, since
the NP does not provide enough surface area for long chains to be flattened against;
the same picture emerged also in previous MD studies of the same CG
system~\cite{Muller-Plathe:12}.
A different trend is observed for grafted chains (Fig.~\ref{fig:distrib}c):
in this case, $R_{gg}$ increases with $L_f$ for chains in close contact to the NP surface.
This trend indicates that for high $L_f$, grafted chains assume stretched configurations,
since they are pushed toward the NP surface by the surrounding free chains. The rise of
$R_{gg}$ for $d_{cm} \approx 1$ is then followed by a decay, with a minimum observed at
$d_{cm} \approx 2$, which is in turn followed by a new rise for higher distances.
Hence, in comparison with
free chains, the conformations of grafted chains appear more dependent on $d_{cm}$,
this dependence becoming particularly enhanced for $L_f=2400$.

Beside the elongation of free and grafted chains, we also investigate their orientation
as a function of their distance from the NP surface. We use
the second Legendre polynomial $P_2(r)$ defined as:
\begin{equation}
P_2(r)=\frac{1}{2} \langle 3 {\rm {cos}}^2\theta(r)-1 \rangle
\end{equation}
where $\theta(r)$ is the angle between the NP-surface normal and some vector of interest.
In Fig.~\ref{fig:orient}a we show a schematic representation of such an angle when the vector
of interest joints nearest neighbours (1-2) or next-nearest neighbours (1-3).
Results for $P_2(r)$ for free and grafted chains are reported in panels (b) and (c) of
Fig.~\ref{fig:orient} for all 1-2 and 1-3 segment pairs.
The figure shows that the distributions are independent of the length of the
free chains, since the 1-2 and 1-3 inter-bead vectors are governed by local
length scales~\cite{Muller-Plathe:12,Ndoro:11}, not by the chain size. Segment pairs
which get very close to the NP surface ($d_{cm}<3$ nm) orient parallel to it
(and perpendicular to the normal, whence $P_2(r)<0$). Above this distance they are
randomly oriented ($P_2(r)=0$). The transition is sharper for 1-2 vectors
(Fig.~\ref{fig:orient}b) than for 1-3 vectors (Fig.~\ref{fig:orient}c).
For grafted chains we find an intermediate region (around 3-4 nm) with a very small
preference of perpendicular orientation with respect to the surface. These vectors
connect one monomer, which is adsorbed on the surface, to one in the layer above.
Therefore the vector connecting them points away from the surface. These calculations
reproduce very well the findings of full MD simulations~\cite{Muller-Plathe:12,Ndoro:11}.

In summary,
comparing our results with previous simulation studies with a full particle-particle
potential~\cite{Ndoro:11,Muller-Plathe:12}, we observe that all essential
features characterizing the nanocomposite are well reproduced within the hybrid particle-field
approach. In addition, the possibility to fully relax the systems even for
long polymer chains paves the
way for the investigation of equilibrium properties, very hard to describe with the traditional
simulation approaches. In particular, the behavior of one of such properties, the PMF,
is the object of a detailed investigation in the following subsections.

\subsection{Two-nanoparticles interactions}
\begin{table*}[t!]
\begin{center}
\caption{Nanocomposite systems considered in the calculation of the two-body 
PMF. Each case is labeled by a different symbol. 
The box lengths are $L_x=22$ nm, $L_y=L_z=12.5$ nm. 
}\label{tab:2PMF}
\begin{tabular*}{0.90\textwidth}{@{\extracolsep{\fill}}ccccccc}
\hline
\hline
& Grafting density &  Grafted chains  
& Grafted chains &  Free chains &
 Free chains \\
& ($\rho_g$) &  ($N_g$)  
& length ($L_g$) &  ($N_f$) &
 length ($L_f$) \\
\hline
& 0  &  0 &  - &  1044 &  20 \\
& 0  &  0 &  - &  104 &  200 \\
& 0.5 ({\bf +})  &  25 &  80 &  854 &  20 \\
& 1.0 ($\ast$)  &  50 &  20 &  944 &  20 \\
& 1.0 ($\times$)  &  50 &  80 &  644 &  20 \\
& 0.5 ($\circ$)  &  25 &  80 &  17 &  1000 \\
& 1.0  ($\square$)  &  50 &  20 &  19 &  1000 \\
& 1.0 ($\triangle$)  &  50 &  80 &  13 &  1000 \\
& 0.04($\bullet$)  &  2 &  40 &  1028 &  20 \\
& 0.1 ($\blacksquare$) &  5 &  80 &  995 &  20 \\
& 0.4 ($\blacktriangle$) &  20 &  20 &  99 &  200 \\
\hline
\end{tabular*}
\end{center}
\end{table*}

The total collection of the investigated systems for the calculation of 
the two-body PMF between a pair of ungrafted or grafted NPs is reported in Tab.~\ref{tab:2PMF}: 
for clearly identifying the various cases, we have labeled each system with a different symbol.
Note that the simulation box has been taken rectangular,
with the long side oriented along
the axis connecting the centers of mass of the two NPs.
In order to obtain the full interaction, we have first calculated the
interaction between the cores of the two NPs
by performing all-atom MD simulations
in the vacuum and comparing the results with an analytical theory developed by
Hamaker~\cite{Hamaker:37} (see the {\bf Supporting Information}), following the procedure described and validated
in Ref.~\cite{Munao:18}. 

\begin{figure}[t!]
\begin{center}
\begin{tabular}{cc}
\hspace{-1.2cm}
\includegraphics[width=4.8cm,angle=0]{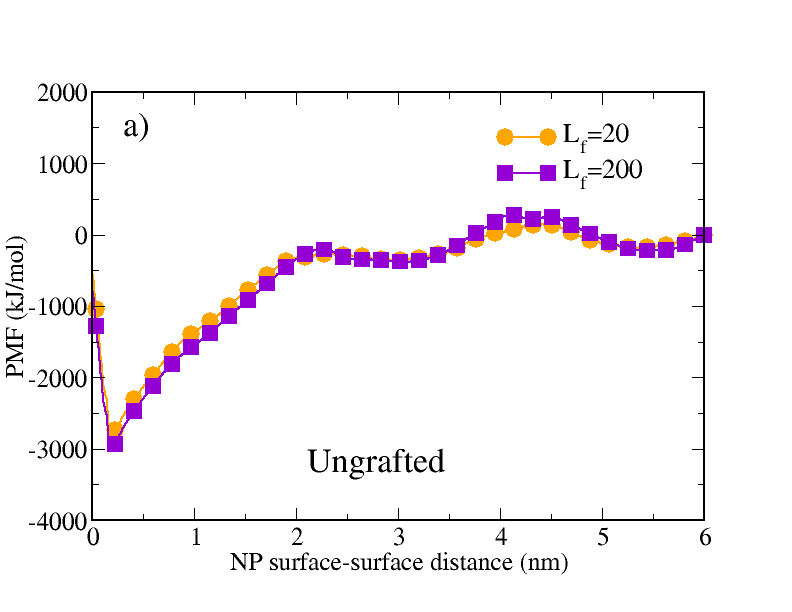} 
\includegraphics[width=4.3cm,angle=0]{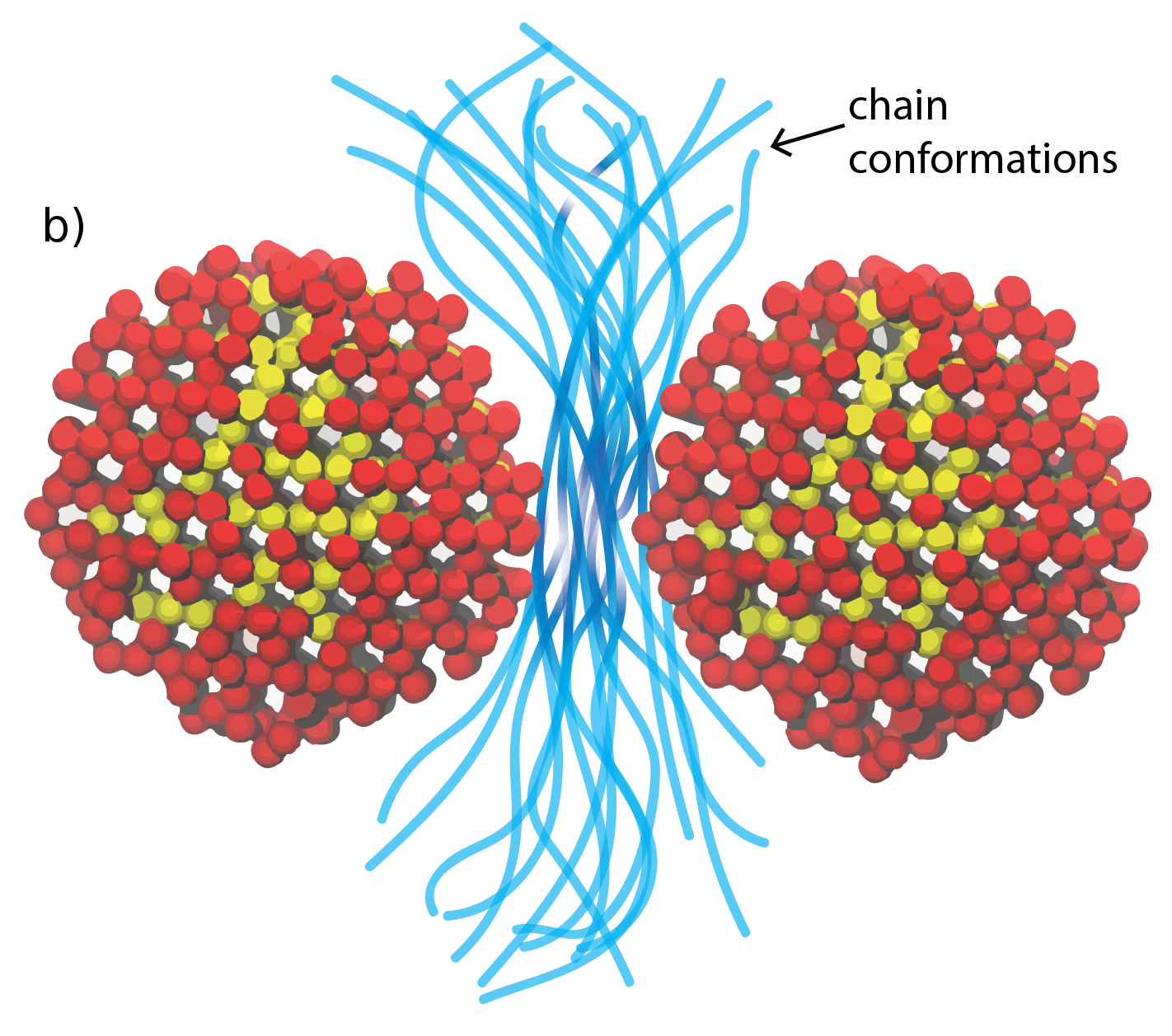} \\
\hspace{-1.2cm}
\includegraphics[width=4.8cm,angle=0]{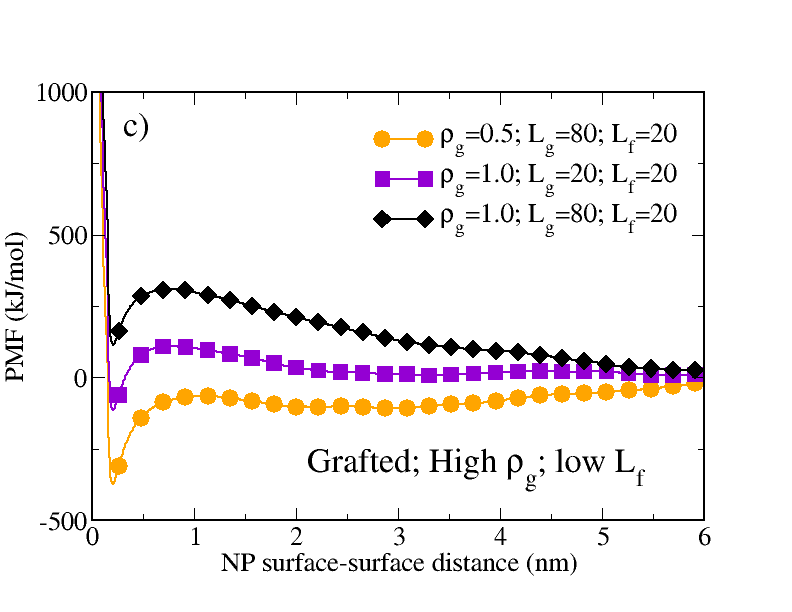}
\includegraphics[width=4.3cm,angle=0]{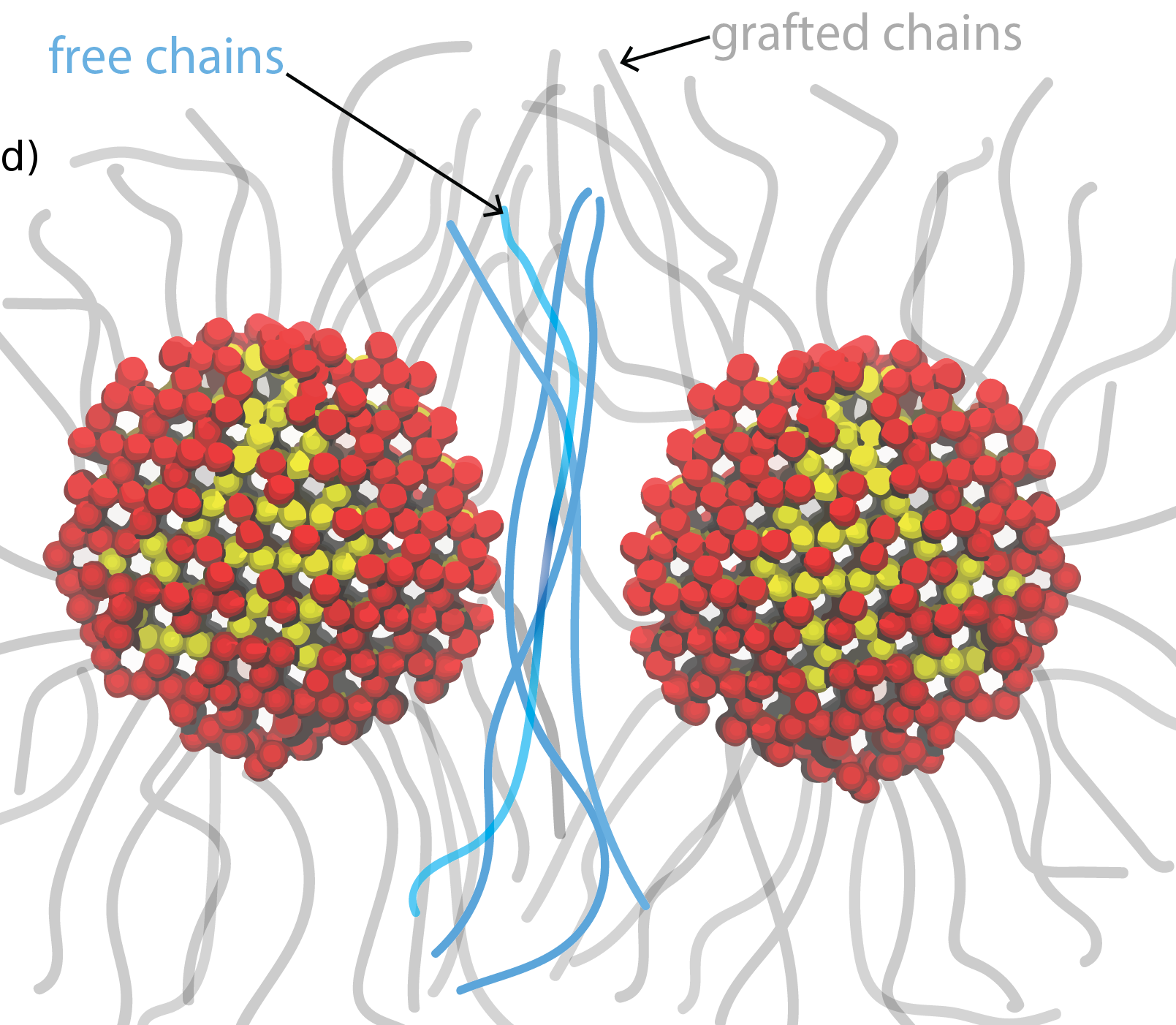} \\
\hspace{-0.6cm}
\includegraphics[width=4.8cm,angle=0]{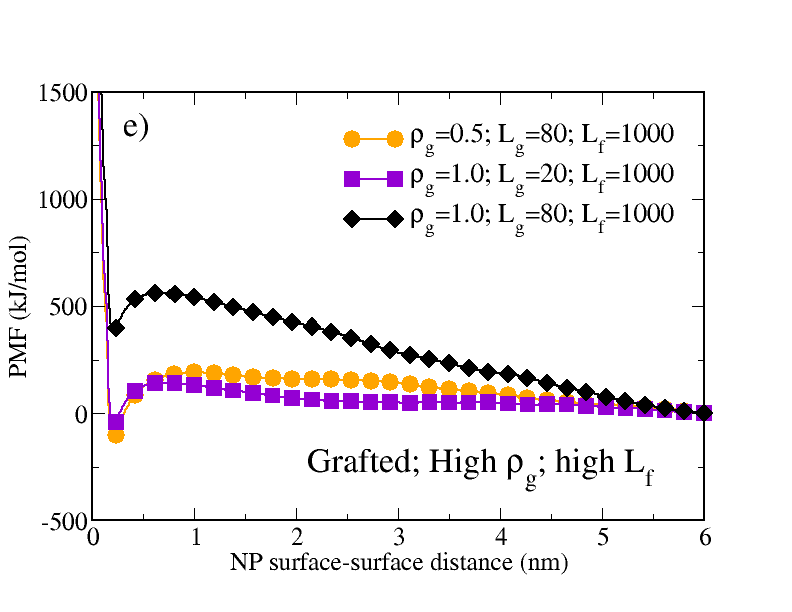} 
\includegraphics[width=4.8cm,angle=0]{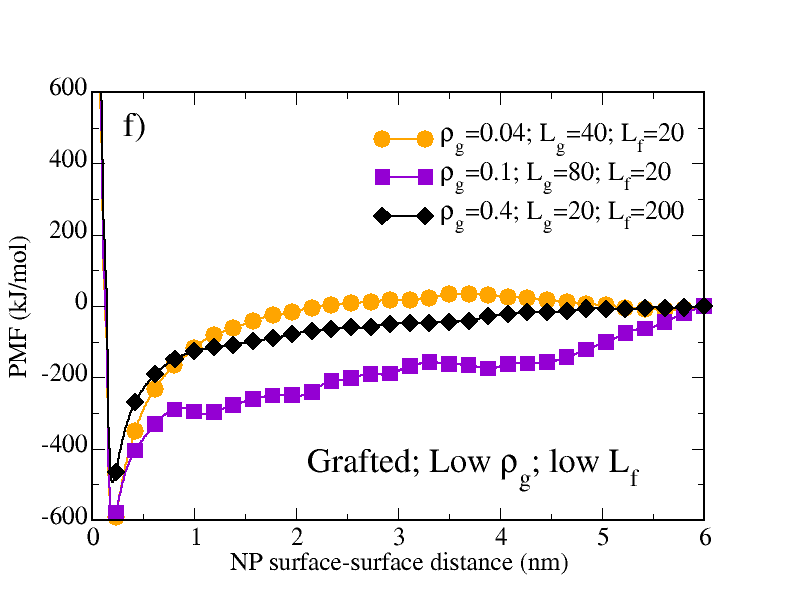}
\end{tabular}
\caption{Behavior of the two-body PMF between ungrafted and grafted
NPs as a function of their mutual distance.
Values of $L_f$, $\rho_g$ and $L_g$ are reported in the legends. 
Cartoons showing confinement of a polymer chain conformation between two ungrafted or grafted NPs
are given in panels (b) and (d): grafted and free chains are reproduced in grey and blue, respectively,
whereas other chains are not shown.
}
\label{fig:PMF-2nak}
\end{center}
\end{figure}
The two-body PMF between
ungrafted and grafted NPs included in a PS matrix of variable chain length
are reported in Fig.~\ref{fig:PMF-2nak}.
In the ungrafted case (panel a) the interaction 
is strongly attractive and independent on the chain length: this behavior 
is in agreement with previous DFT~\cite{Yeth:11,Patel:04} 
and PRISM~\cite{Hooper:04} studies of the PMF in simple CG models 
and can be understood
in terms of chains confinement: in panel (b) we show a cartoon representing 
several configurations of a single
chain confined between the NPs for a interparticle separation of 1 nm. 
\begin{figure}[t!]
\begin{center}
\includegraphics[width=8.5cm,angle=0]{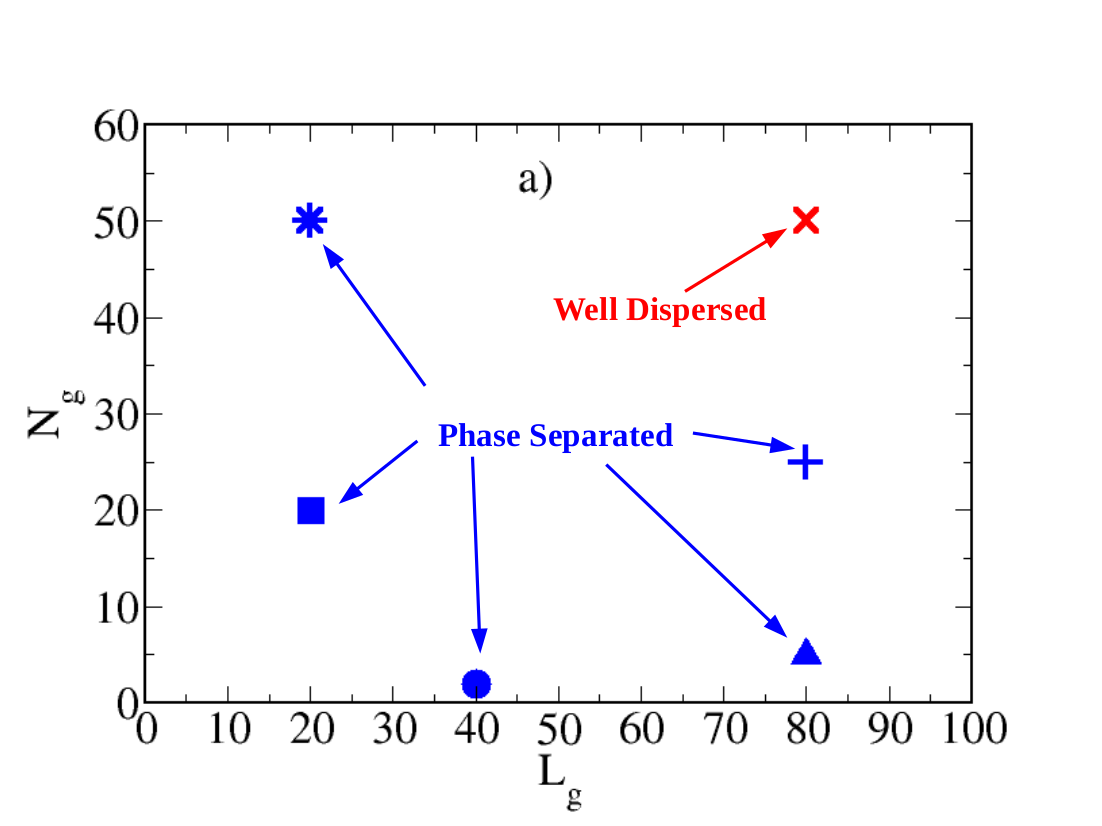} \\
\includegraphics[width=8.5cm,angle=0]{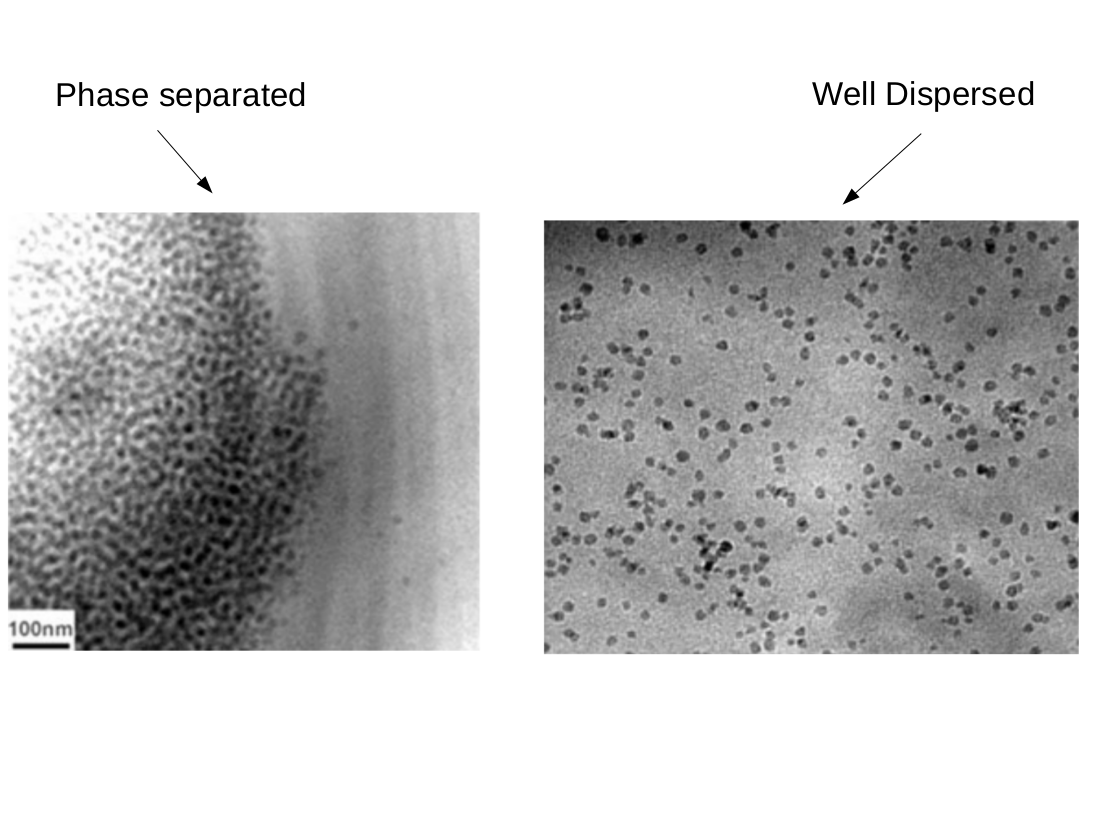} \\
\vspace{-0.9cm}
b)
\vspace{0.5cm}
\caption{Panel (a): schematic phase behavior for grafted NPs embedded in a PS matrix, drawn by considering
the two-body PMF.  
Different colors are used for well dispersed ($B_2 >0$) and phase separated ($B_2 < 0$) regions.
Panel (b): trasmission electron microscopy images of the corresponding structures, 
redrawn from Lan et al.~\cite{Lan:07}. 
The scale bar is indicated in the first image of the panel.}
\label{fig:phase}
\end{center}
\end{figure}
As visible, in the confinement zone the chain is considerably compressed, whereas outside this zone
it is able to explore a larger variety of geometrical configurations.
Therefore, chains can be arranged between the two NPs only by assuming extended conformations: 
the high entropic penalty of this conformation  
gives rise to a strong attraction between the NPs. Upon increasing
the interparticle distance, a progressively higher number of chains can be confined 
in a more coiled states and the attraction diminishes. 
It is also worth noting that
the PMF oscillates until reaching the zero value
with a period corresponding to the radius of a NP.
This is in agreement with previous theoretical~\cite{Hooper:04,Hooper:05,Yeth:11} and
simulation~\cite{Bedrov:03} studies, where it was found that
oscillations develop when the polymer chains are
confined between the NPs and the regions of perturbed polymer density around each
NPs begin to overlap. More specifically, the oscillatory behavior indicates a 
correlation between the structure 
of the polymer in the interparticle region and matrix-induced 
NP-NP interactions.
For particles in close contact, oscillations can not develop
and a monotonic decay toward strongly negative values of the PMF is 
found, as expected for high NP-PS repulsive interactions. 

A different scenario is observed for the grafted cases: for high grafting 
densities and low molecular weigths of the free chains (panel c) the PMF is
mainly repulsive, such repulsion becoming stronger upon increasing $\rho_g$ or $L_g$.
The cartoon reported in panel (d) clarifies the origin of the NP-NP repulsion: in comparison with 
the ungrafted case, now a large number of grafted chains are forced to be confined between the two NPs; 
even if a small number of free chains can still be confined,
the available space contains especially the grafted ones, that push the two NPs away from each 
other, hence generating the repulsion between them.  

If the free chain length increases (panel e), the repulsion between the NPs increases in turn, 
and this is especially clear
for $\rho_g=0.5$ chains/nm$^2$ and $L_g=80$: in fact the minimum of the total PMF, previously found
at $\simeq$-380 kJ/mol, is now observed at $\simeq$-140 kJ/mol. As a consequence,
the total PMF for $\rho_g=0.5$ chains/nm$^2$ and $L_g=80$ and for $\rho_g=1.0$ chains/nm$^2$ 
and $L_g=20$ are now almost coincident. 
The increase of the short-range repulsion between a pair of 
grafted NPs upon increasing $\rho_g$ and $L_f$ is also in agreement with
previous simulation studies of bead-spring models~\cite{Smith:09,Meng:12}. 
The analysis is completed by exploring cases corresponding to low grafting densities and
free chain lengths (panel f): unlike what has been observed so far, the PMF shows a 
pronounced attractive well, keeping
negative values for almost the entire range of interparticle distances. The attraction
strength increases upon decreasing $\rho_g$, with the PMF monotonically going to zero,
except for $\rho_g=0.04$ chains/nm$^2$, where a maximum is observed for an interparticle
distance of 4 nm. We infer that such a feature could be reminiscent of the oscillating behavior
observed for ungrafted NPs, since the grafting density is quite low.
For $\rho_g=0.4$ we have also verified that upon decreasing $L_f$ from 200 to 20, no
qualitative difference in the PMF are observed (see the {\bf Supporting Information}). 

By collecting these results, it 
emerges that the
molecular weight of the free chains does not significantly influence the PMF.
On the other hand,
the grafting density plays a critical role in determining the behavior of the effective
interactions between grafted NPs: for high $\rho_g$ the two-body PMF is always repulsive, 
whereas for low $\rho_g$ it becomes attractive. 
As a consequence, a system comprising several NPs in a PS
melt should be well dispersed in the first case and phase-separated in the second case. 
This assumption can be verified by calculating the second virial coefficient $B_2$ from the PMF, since
positive values of $B_2$ identify regions of the phase diagram where repulsive contributions are dominant,
whereas its negative values indicate that attractive interactions prevail. The general definition of the
second virial coefficient for a potential without an angular dependence can be written as~\cite{McQuarrie:76}:
\begin{equation}\label{eq:b2}
B_2=2\pi \int_0^{\infty} (1-e^{-\beta U(r)}) r^2 dr
\end{equation}
where $U(r)$ is the PMF, $\beta=k_B T$ and $r$ is the interparticle separation. In our case the interval of
integration $[0,\infty]$ is replaced by $[r_{min},r_{max}]$ according to the definition of Eq.~\ref{eq:PMF}. 
By performing the calculation of $B_2$, keeping into account the effects due to $\rho_g$ and $L_g$ 
and following a prescription reported
in Ref.~\cite{Akcora:09}, it is possible to draw 
a schematic representation of the phase behavior of the nanocomposite as obtained from the two-body PMF and
reported in panel (a) of Fig.~\ref{fig:phase}. Experimentally obtained images of well dispersed and phase
separated conditions (redrawn from Lan et al.~\cite{Lan:07}) are shown in panel (b). 
It is noteworthy to compare such findings with simulation~\cite{Akcora:09} and 
experimental~\cite{Lan:07,Chevigny:11,Sunday:12,Kumar:13} 
data on the collective behavior of grafted NPs embedded in a PS matrix: we first note that,
unlike what we find, it is experimentally observed that for high molecular weights of the free chains
a phase separation is detected. However, it must be pointed out that all data reported in 
Ref.~\cite{Lan:07,Chevigny:11,Sunday:12,Kumar:13}
refer to NPs of bigger size than that investigated here. Previous experimental~\cite{Archer:12} and 
simulation~\cite{Hall:10} studies on the effects of the NP 
size on the PMF have effectively proved that for small NPs (when the radius of gyration of the grafted chains
is comparable to the NP radius) the effective interactions between NPs are mainly repulsive. In particular,
a DFT study~\cite{Hall:12} has established that in good solvent conditions the PMF
betwen two grafted NPs is always repulsive. 
The results obtained in these studies explain the repulsive
behavior of the two-body PMF observed in our simulations. 
\begin{table*}[t!]
\begin{center}
\caption{Nanocomposite systems considered in the calculation of the three-body 
PMF. Note that all grafted cases correspond to those reported in Tab.~\ref{tab:2PMF}, except for the values
of $N_f$. The box length is $L_b=22$ nm.}
\label{tab:3PMF}
\begin{tabular*}{0.90\textwidth}{@{\extracolsep{\fill}}cccccc}
\hline
\hline
& Grafting density &  Grafted chains  & Grafted chains &  Free chains & Free chains \\
& ($\rho_g$) &  ($N_g$)  & length ($L_g$) &  ($N_f$) & length ($L_f$) \\
\hline
& 0  &  0 &  - &  2796 &  20 \\
& 0.5   &  25 &  80 &  2527 &  20 \\
& 1.0   &  50 &  20 &  2662 &  20 \\
& 1.0   &  50 &  80 &  2194 &  20 \\
& 0.5   &  25 &  80 &  50 &  1000 \\
& 1.0   &  50 &  20 &  53 &  1000 \\
& 1.0   &  50 &  80 &  44 &  1000 \\
& 0.04  &  2 &  40 &  2792 &  20 \\
& 0.1   &  5 &  80 &  2775 &  20 \\
& 0.4   &  20 &  20 &  274 &  200 \\
\hline
\end{tabular*}
\end{center}
\end{table*}
%
The comparison with simulation and experimental morphology diagrams reported in Refs.~\cite{Akcora:09,Kumar:13} 
shows a futher
interesting feature: in those diagrams, for low grafting densities the NPs are found to self-assemble in
complex nanostructures, like strings, connected sheets and small clusters. The appearance of a given 
structure strongly depends on a proper combination of $\rho_g$, $L_g$ and $L_f$. According to 
the behavior of the two-body PMF at  
low grafting densities, reported in Fig.~\ref{fig:PMF-2nak}f, 
it emerges that the resulting NP-NP
interactions are purely attractive for all the interparticle distances and then  
cannot
reproduce the appearance of complex structures. 
This is due to the subtle competition between
attractive and repulsive contributions (rather than a clear prevalence of one of the two effects)
which takes places for low grafting densities, where these
structures appear.
The crucial role played by the grafting density in determining the structure of 
nanoscale aggregates has been recently stated also in Ref.~\cite{Bachhar:17}.
On the other hand, a proper characterization of this region of the morphology diagram is important,
due to the variety of possible technological applications connected to the appearance of
nanostructures. Multi-body effects can play a significant role for a proper
dealing of the competition between attractive and repulsive effects, as shown in the
next subsection, focused on the three-body contribution to the global PMF.

\subsection{Three-nanoparticles interactions}
\begin{figure*}[t!]
\begin{center}
\includegraphics[width=9.0cm,angle=0]{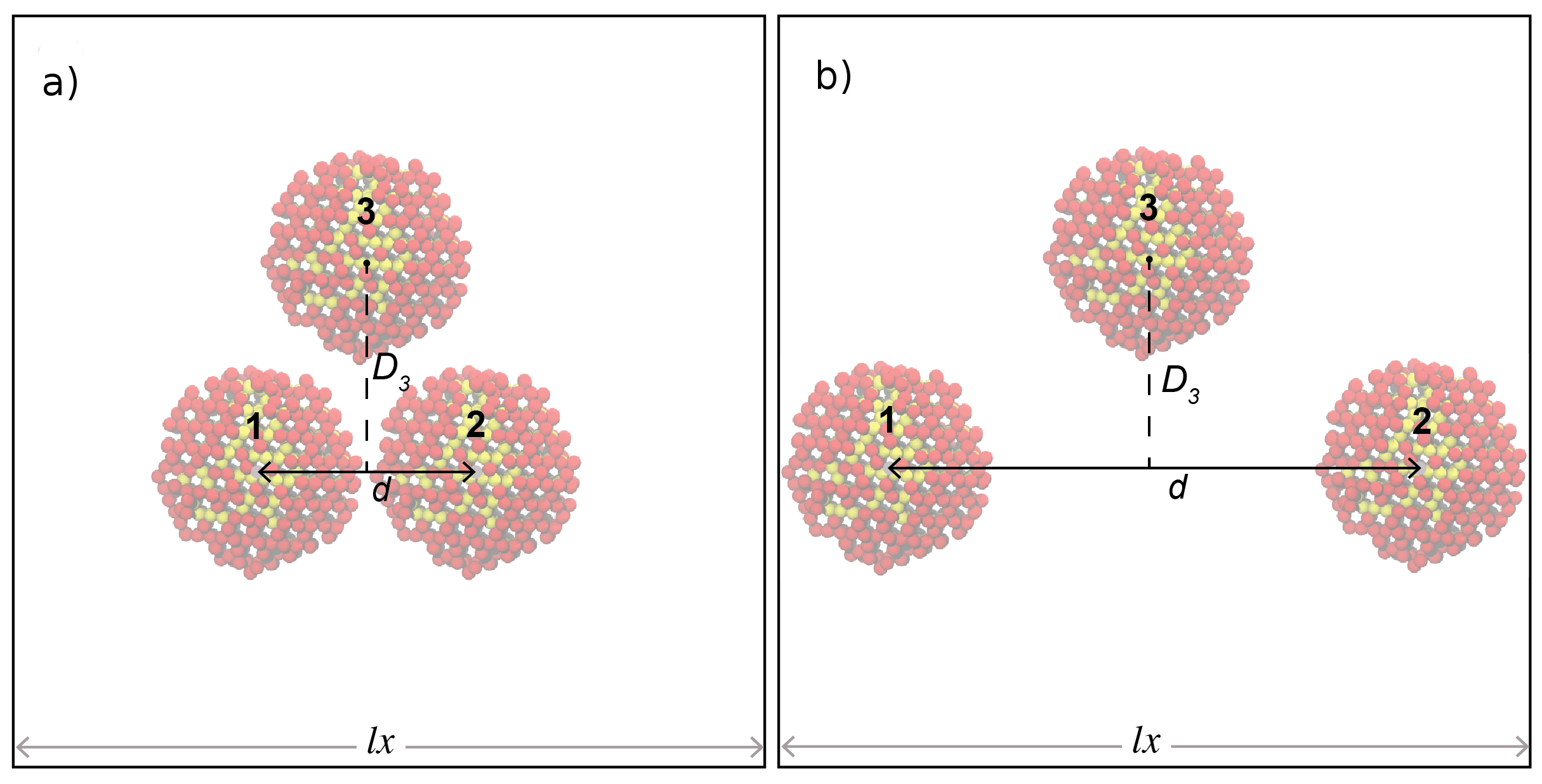} \qquad
\includegraphics[width=6.0cm,angle=0]{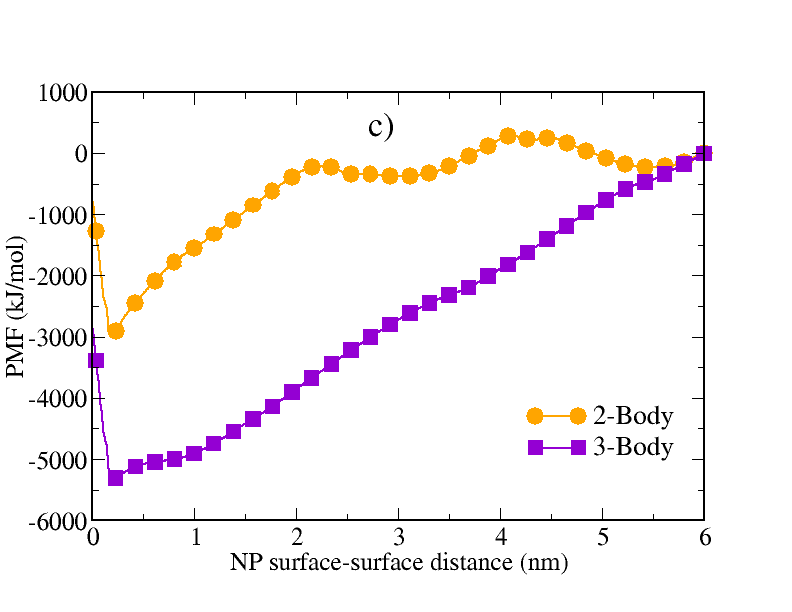} \\
\includegraphics[width=10.0cm,angle=0]{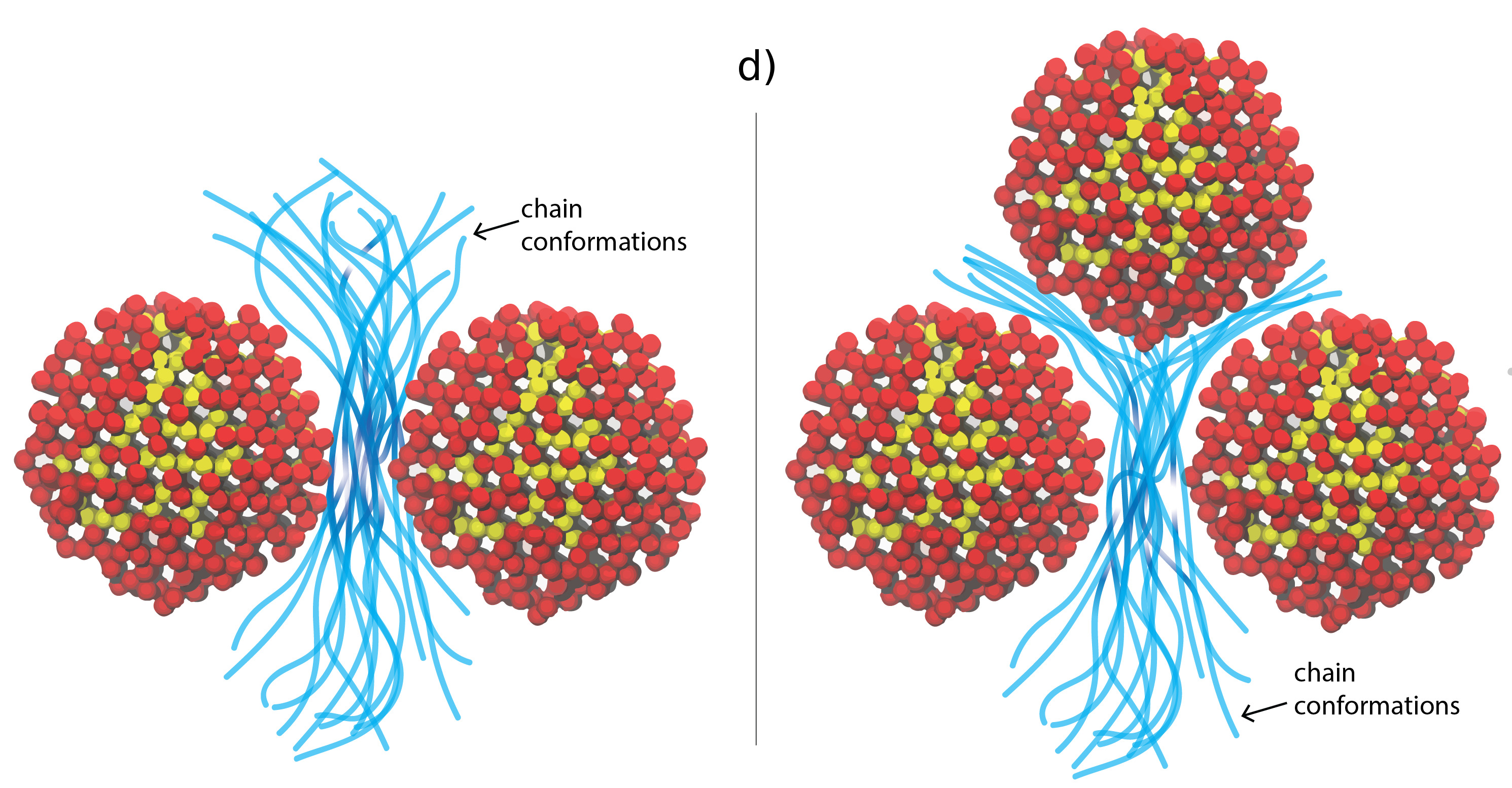} 
\caption{Panels (a), (b): schematic representation of three ungrafted NPs in a PS melt, 
suited for the
calculation of the three-body PMF: the distance $D_3$ between the
center of the NP 3 and the center of the 1-2 vector is 4 nm. 
Panel (c): comparison between two-body and three-body PMF between
ungrafted NPs with $L_f=20$ as a function of their mutual distance for $D_3=4$ nm.
Panel (d): cartoon showing the restricted conformational freedom of a polymer chain confined
between three NPs (in comparison with the two-body case); 
other chains are not shown.}
\label{fig:PMF-3nak}
\end{center}
\end{figure*}
The total collection of the investigated systems for the calculation of
the three-body PMF 
is reported in Tab.~\ref{tab:3PMF} and
the schematic procedure for its calculation is given in panels (a), (b)
of Fig.~\ref{fig:PMF-3nak}: the systems are now made of a variable
number of PS chains plus three NPs (for clarity labeled 1, 2 and 3),
whose mutual distances are 
kept fixed. These NPs are placed in a T-shaped arrangement with the
horizontal bar being the 1-2 vector, and the vertical bar being the vector
between the 1-2 midpoint and particle 3 at distance $D_3$. The contribution due to the
presence of particle 3 is calculated by computing
the total forces acting on the centers of mass of particles 1 and 2. The two-body PMF corresponds 
to a particular case of the three-body PMF where $D_3=\infty$; due to the particular 
arrangement of the three NPs, the shape of the
simulation box is now cubic, with $L_b=22$ nm. 
The comparison between two and three-body 
PMF between ungrafted NPs embedded in a PS matrix constituted by chains of 20 beads 
and for $D_3=4$ nm is reported in panel (c) of Fig.~\ref{fig:PMF-3nak}.
It emerges that the addition a third NP in the simulation box 
has a remarkable effect on the resulting interaction, since 
the three-body PMF is considerably more attractive than the two-body
counterpart.
Furthermore, oscillations observed for the two-body case are now suppressed, this indicating
that the presence of the third NP breaks the previously described mechanism of association,
giving rise to a PMF without intermediate minima that monotonically goes to zero 
for large interparticle
distances. The increase of the attraction can be understood by looking at the cartoon
reported in Fig.~\ref{fig:PMF-3nak}d: in comparison with the two-body case, 
a chain confined between three NPs is now even more compressed, 
and the accessible space just outside
the confinement zone is further reduced. Hence, chains have less conformational freedom between the NPs,
this generating a strong attraction between them.

\begin{figure*}[t!]
\begin{center}
\begin{tabular}{ccc}
\includegraphics[width=5.5cm,angle=0]{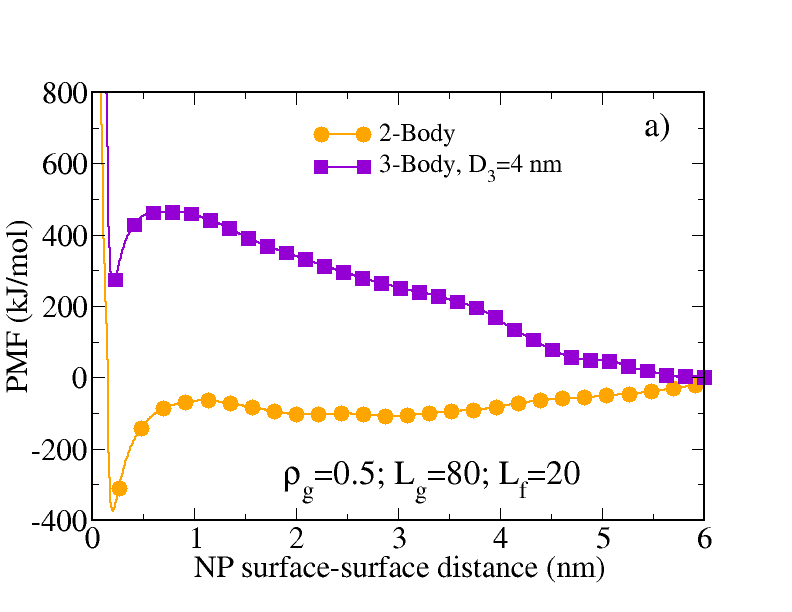} 
\includegraphics[width=5.5cm,angle=0]{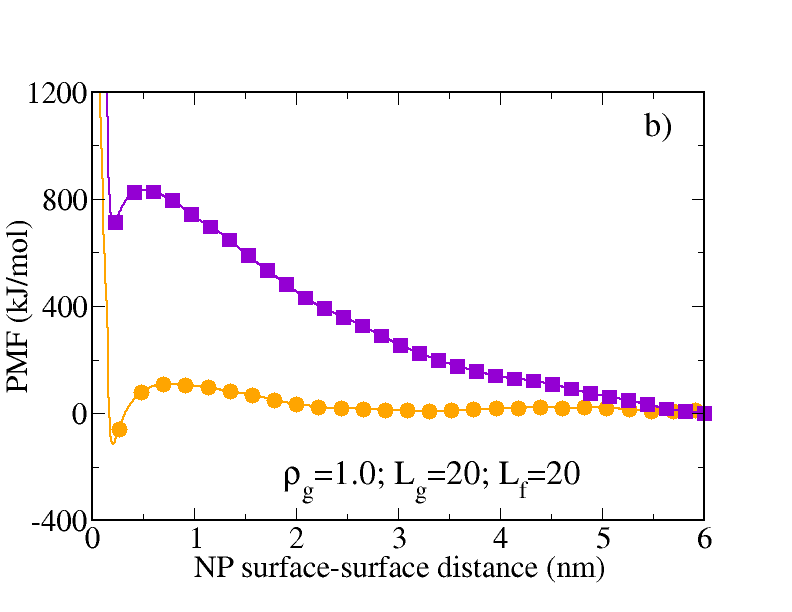} 
\includegraphics[width=5.5cm,angle=0]{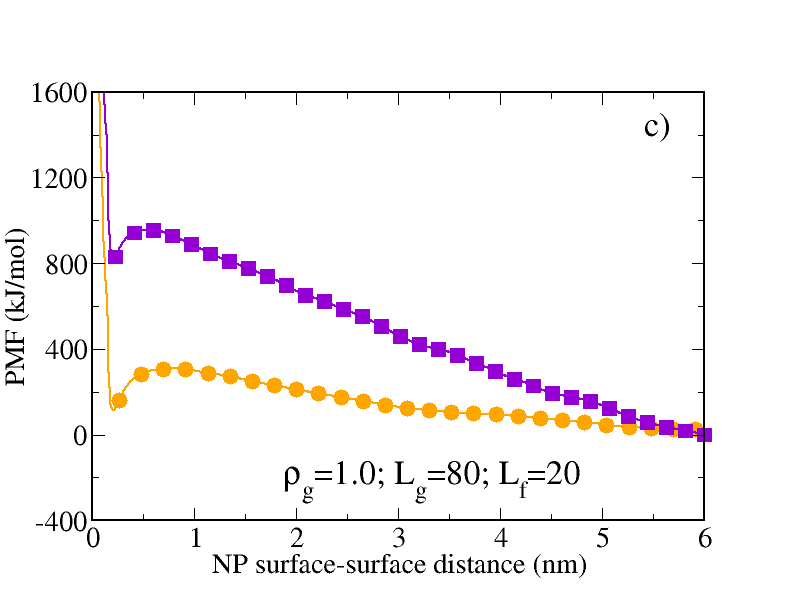} \\
\includegraphics[width=5.5cm,angle=0]{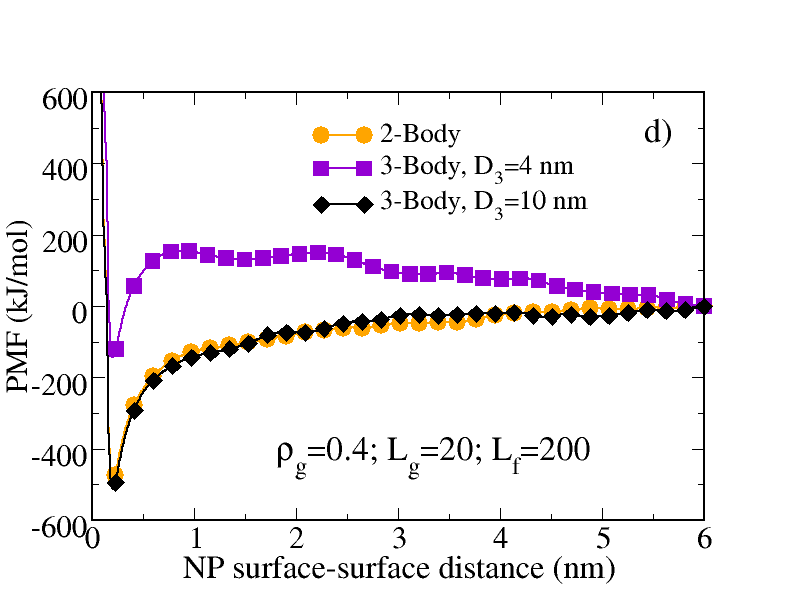} 
\includegraphics[width=5.5cm,angle=0]{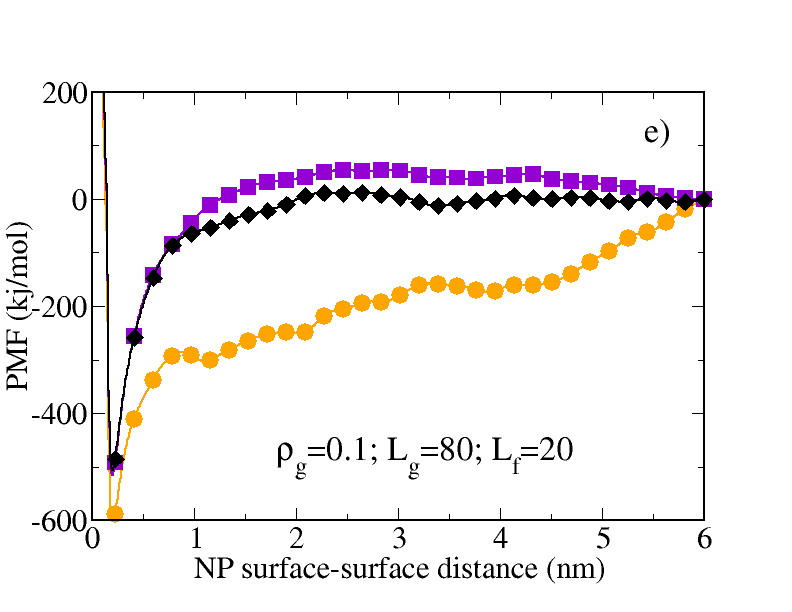} 
\includegraphics[width=5.5cm,angle=0]{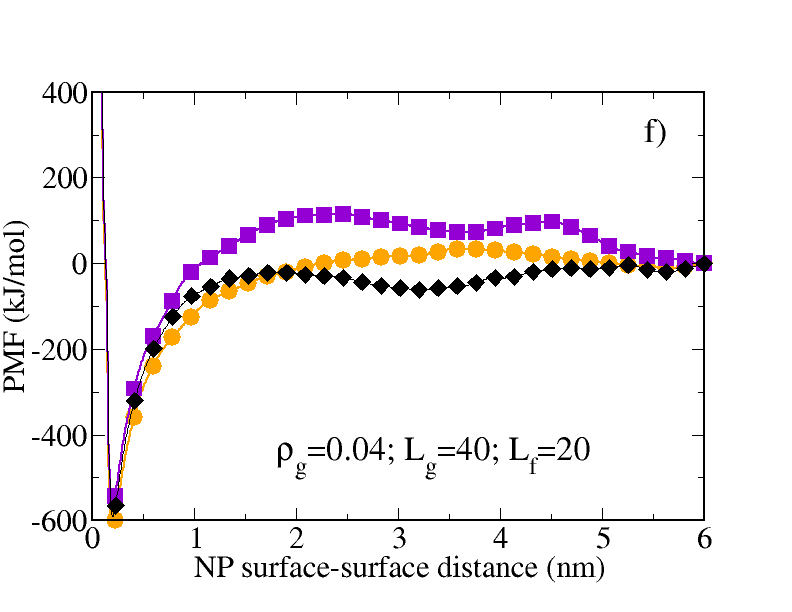} \\
\end{tabular}
\caption{Comparison between two-body and three-body PMF between
grafted NPs as a function of their mutual distance. 
In the top panels cases corresponding to $D_3=4$ nm and high grafting densities are shown.
In the bottom panels we report cases corresponding to $D_3=4$ nm, $D_3=10$ nm and low grafting
densities. Specific values of $\rho_g$, $L_g$ and $L_f$ are given in the legends.
}
\label{fig:PMF-3b-20}
\end{center}
\end{figure*}

The calculation of the three-body contribution among grafted NPs 
follows the same lines considered for the ungrafted cases.
Two and three-body PMF between grafted NPs 
are compared in Fig.~\ref{fig:PMF-3b-20}: in particular, 
in the top panels cases corresponding to high grafting density, low free chain lengths and
$D_3=4$ nm are considered. In these conditions,
the three-body PMF is systematically more repulsive than 
the two-body counterpart,
a behavior opposite to that observed
for ungrafted NPs. This is due to the increase of the steric repulsion between grafted chains, which
becomes bigger and bigger the closer the NPs are. 
The emerging picture is compatible with a system where NPs are in a well dispersed condition.
The same scenario holds if the free chain lengths increases till to 1000 beads in a single
chain (see the {\bf Supporting Information}). 
From a physical point of view, for high grafting densities, the addition of a third particle
in the calculation of effective interactions does not change the scenario observed in the simpler two-body 
case, but strengthens the repulsive behavior of the resulting PMF, in particular for small interparticle
distances. A similar effect has been noticed also in a previous DFT study of the three-body interactions in
polymer nanocomposites~\cite{Yeth:11} and it has been related to the existence of a small region of 
polymer confined between the three NPs, which gives rise to large contribution to the PMF. The authors
hence concluded that three-body interactions cannot be neglected in a proper description of the
PMF between NPs in polymer melts, since they bring a significant contribution to the total interaction.
\begin{figure}[t!]
\begin{center}
\includegraphics[width=8.5cm,angle=0]{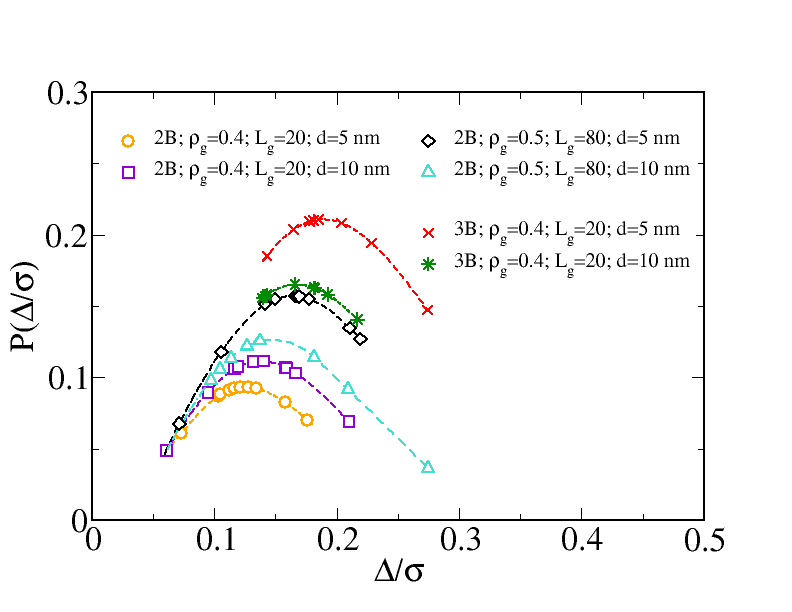}
\caption{Distribution of $\Delta$ values for different grafting densities, grafted chain lengths and NP 
center-center separations (indicated in the legends). Both two (2B) and three-body (3B) 
cases are considered. For the three-body PMF we have chosen $D_3=4$ nm.
The interparticle distances are normalized by the NP diameter $\sigma$.} 
\label{fig:delta}
\end{center}
\end{figure}
\begin{figure*}[t!]
\begin{center}
\includegraphics[width=15.0cm,angle=0]{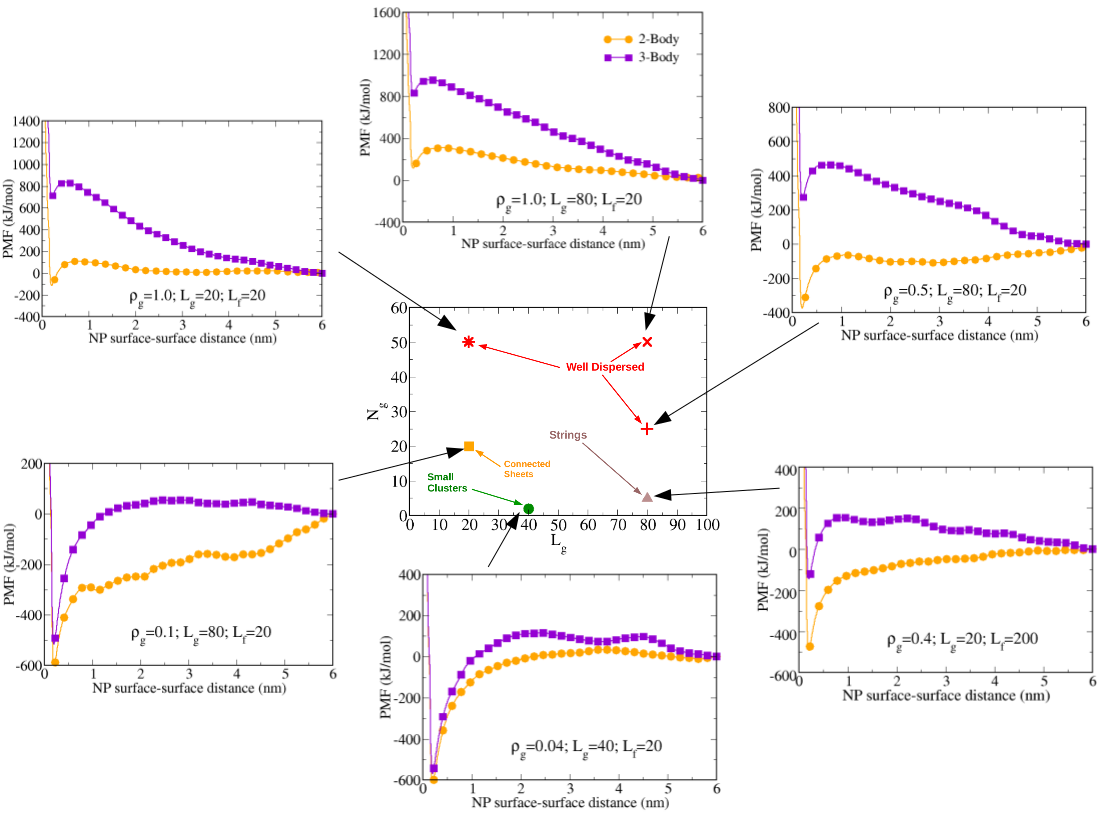} \\
\includegraphics[width=4.0cm,angle=0]{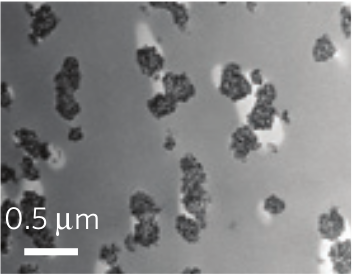} \quad
\includegraphics[width=4.0cm,angle=0]{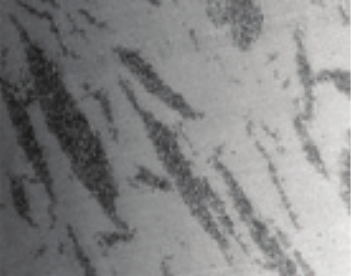} \quad
\includegraphics[width=4.0cm,angle=0]{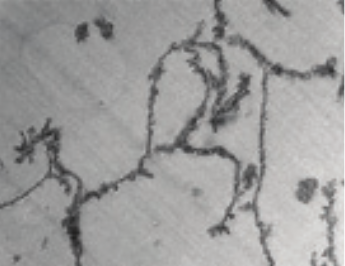} \\
\hspace{-1.0cm} Small Clusters \hspace{1.8cm} Connected Sheets \hspace{2.0cm} Strings \\
\caption{Top: schematic morphology diagram for grafted NPs embedded in a PS matrix, drawn by considering
the three-body PMF for $D_3=4$ nm and calculating the second virial coefficient.
Different colors are used for well dispersed, strings, connected sheets and
small clusters regions. 
The behaviors of two-body and three-body PMF are also reported for enhancing their relation
to the morphology diagram.
Bottom: trasmission electron microscopy images of the corresponding self-assembled structures,
redrawn from Akcora et al.~\cite{Akcora:09}.
The scale bar is indicated in the first image of the panel.
} 
\label{fig:Kumar-tot}
\end{center}
\end{figure*}

The most interesting situation is that corresponding to low grafting densities (bottom panels
of Fig.~\ref{fig:PMF-3b-20}), where the two-body
PMF is attractive, thus suggesting aggregation between NPs.
The comparison between two and three-body PMF under these conditions is done 
for $D_3=4$ and $D_3=10$ nm, in order to take into account also the
effects due to the range of the three-body potential.
For $D_3=4$ nm all three-body PMF
show the existence of a short-range attraction followed by a repulsion, the strength of the two interactions
being strictly dependent on the specific cases considered. The contemporary presence of attractive and
repulsive contributions in the total interactions is usually related to the onset of self-assembled
complex structures~\cite{Stradner}.
More specifically, for $\rho_g=0.4$ chains/nm$^2$ and $D_3=4$ nm (panel d)
the three-body PMF
has now a shallow attractive minimum followed by a slightly repulsive flat region before going to zero.
A comparison with a simulation morphology diagram reported in Ref.~\cite{Akcora:09} suggests that
strings should appear under these conditions.
For $\rho_g=0.1$ chains/nm$^2$ and $D_3=4$ nm (panel e) 
the three-body PMF shows a quite deep minimum (similarly to the
two-body case), before attaining a plateau around zero already for short-range distances.
It has been shown that very short-range attractive potential can give rise to planar
structures~\cite{Preisler:14} resembling the connected sheets reported in Refs.~\cite{Akcora:09,Kumar:13}. 
However this behavior was
found for anisotropic potentials: whether this can happen also for the PMF shown in
Fig.~\ref{fig:PMF-3b-20}e, is not straigthforward to infer.
Finally, for $\rho_g=0.04$ chains/nm$^2$ and 
$D_3=4$ nm (panel f) the three-body PMF still shows a deep sharp minimum, followed
by a repulsive tail persisting even to intermediate NP-NP separation and becoming zero only
for distances close to 6 nm. The simultaneous presence of a short-range attraction and a 
long-range repulsion is usually related to the appearance of clusters~\cite{Attr_Rep}, 
as expected also from previous simulation and experimental morphology diagrams~\cite{Akcora:09,Kumar:13}. 
The picture is completed by analyzing the range dependence of the three-body PMF:  
for $\rho_g=0.4$ chains/nm$^2$ (panel d)
the three-body PMF strongly depends on the distance $D_3$, since for $D_3=10$ nm we already recover the
behavior of the two-body PMF. Hence, under these conditions the three-body
effects are significant only for short interparticle distances, as can be expected if strings 
effectively appear in the system.
Conversely, for $\rho_g=0.1$ chains/nm$^2$ (panel e) the three-body effects are long-ranged, since for
$D_3=10$ nm the PMF still shows the same behavior observed for $D_3=4$ nm: this can be
compatible with the presence of sheets, since in order to bond many NPs in a planar configuration,
the three-body contribution must persist even for large
interparticle distances. Finally, for $\rho_g=0.04$ chains/nm$^2$ (panel f) the PMF for $D_3=10$ nm is again 
similar to the two-body case, hence suggesting that the three-body effects are short-ranged under
these conditions, in agreement with the possible presence of clusters. It may be worth noting that the
short-range or long-range nature of the three-body effects does not monotonically depend on the 
grafting density, but only on the specific system considered.

The dependence of self-assembled structures on multi-body interactions has been highlighted also in previous
numerical studies of solutions containing grafted NPs~\cite{Lane:10}; on the other hand, it also has been 
shown~\cite{Bozorgui:13} that these complex structures can arise due to the anisotropy in the grafted coronas
surrounding the NP cores. Such anisotropy can be investigated by
calculating the asymmetry in the polymer grafting; following the prescription reported in Ref.~\cite{Bozorgui:13}
we have calculated the center of mass $\Delta$ of the grafted chains respect to the center of mass of the NP
to whom they are linked. In Fig.~\ref{fig:delta} we report the probability $P(\Delta)$ of finding different
$\Delta$ values upon varying the grafting density, the grafted chain length and the interparticle distance.
In addition, two and three-body cases (for $\rho_g=0.4$ chains/nm$^2$ and $L_g=20$) 
are assessed against each other. The distributions follow a Gaussian form that reads as: 
\begin{equation}
P(\Delta) \propto \Delta^2 {\rm exp} [-(\Delta^2/\langle \Delta^2 
\rangle)] \,,
\end{equation} 
where $ \langle \Delta^2 \rangle $ is the mean squared value of $\Delta$.
As visible, no peak for $\Delta=0$ is
found, this indicating the existence of an asymmetry in the chain distribution for all cases, as documented also in 
Ref.~\cite{Bozorgui:13}. It is worth noting that for the three-body case and for $L_g=80$ the position of the
peak of $P(\Delta)$ shifts towards low values of $\Delta$ upon increasing the interparticle distances; at the same 
time, also the heigth of the peak decreases. Hence, for low interparticle distances, the grafted coronas are
more warped, which is consistent with the previous description of chain arrangements (see Fig.~\ref{fig:PMF-2nak}    
and Fig.~\ref{fig:PMF-3nak}). The trend is reversed for the two-body case and $L_g=20$, this indicating the
interpenetration between grafted coronas belonging to the two NPs. Overall, the behavior of $P(\Delta)$ indicates
that the asymmetry increases if multi-body interactions are taken into account,  
as expected is nano-structured aggregates onset in the system. 

Summarizing the main outcomes of the three-body PMF it is possible to redraw the schematic
phase behavior reported in Fig.~\ref{fig:phase} 
by calculating the second virial coefficient for the three-body PMF through the Eq.~\ref{eq:b2}. 
However, it is worth noting that the knowledge of $B_2$ provides information only on the repulsive or
attractive nature of the PMF and hence can not identify the specific self-assembled structures. 
As a consequence, the latter can be inferred only by the specific behaviors of the PMF reported in 
Fig.~\ref{fig:PMF-3b-20} and discussed above.
The resulting scheme is shown in Fig.~\ref{fig:Kumar-tot},
along with experimentally obtained images of small clusters, strings and connected sheets 
redrawn from Akcora et al.~\cite{Akcora:09}.
In order to highlight the relation between the PMF and the
different phases, we also report in the same figure 
the behavior of two and three-body PMF (for $D_3=4$ nm) 
at high and low grafting densities. 
As visible, the well dispersed condition, observed in the upper part of the diagram and characterized
by high values of $N_g$ and $L_g$, can be generally well described by the behavior of the
two-body PMF; conversely, for a proper understanding of the nanostructures observed in 
the lower part of the diagram we need to consider the subtle interplay between two-body and
three (or multi) body interactions. 
In this perspective, further refinements of our approach, including
four-body (or higher) contributions to the total PMF are possible, even if it is reasonable to assume 
that higher order corrections become progressively less significant. We deserve explicit calculations 
of four-body contribution to the total interaction to future studies. 

\section{Conclusions}
We have performed a molecular dynamics study of a coarse-grained model of 
Silica-Polystyrene nanocomposites by means of a hybrid particle-field approach.
In this method, the non-bonded interactions among particles are 
decoupled and replaced by a field representation. In this way, it is possible
to drastically reduce the simulation time typically required for properly
relaxing the nanocomposites, opening the way to investigations of 
their local structure
and interfacial properties in detail. Specifically,
spherical Silica nanoparticles (NPs) grafted with Polystyrene chains have been studied, 
aiming to show how the presence of grafted chains influences
the global properties of the nanocomposite. 

The proposed approach has been first validated against standard models previously reported in literature
(based on traditional molecular dynamics simulations) and finding an overall good
agreement with them. In particular, we have verified that the molecular structure 
of polymer chains is well reproduced by the hybrid particle-field approach, 
by documenting the existence of a ``wet-brush-to-dry-brush'' transition, as already observed in
previous experimental~\cite{Chevigny:10} and simulation~\cite{Muller-Plathe:12} works.
In addition, we have checked that even polymer chains with high molecular weight 
can be fully relaxed within the proposed approach.

Then, the potential of mean force (PMF) between a pair of ungrafted or grafted
NPs embedded in the polymer matrix has been calculated. In order to gain
new insight in the knowledge of the effective interactions between the NPs, 
we have calculated, beside the
usual two-body PMF, also the three-body contribution to the total potential. 
To the best of our knowledge,
this is the first time such a term is explicitly taken into account in molecular 
simulations of polymer nanocomposites. 

Upon comparing the PMF between ungrafted and grafted NPs, 
significant differences emerge: while in the first case the PMF is strongly
attractive, for grafted NPs it becomes progressively more repulsive upon increasing
the grafting density. 
The calculation of
three-body PMF between grafted NPs allowed us to make a qualitative comparison with
simulation~\cite{Akcora:09} and experimental~\cite{Kumar:13} diagrams where different morphologies 
(obtained through the self-assembly of many grafted Silica NPs in PS matrices) are reported. 
In particular, we have ascertained how the existence of nanostructures, like strings,
connected sheets and small clusters, can be understood only by taking into account
the subtle interplay between two-body and three (or, more in general, multi) body interactions
in the behavior of PMF. In this perspective, we have proposed our approach and models
for a proper molecular understanding of specific systems, along with the impact of the chemical
nature of such systems on the composite final properties.

\section{Acknowledgements}
The computing resources and the related technical support used for this work
have been provided by CRESCO/ENEAGRID High Performance Computing infrastructure
and its staff~\cite{Cresco}. CRESCO/ENEAGRID High Performance Computing
infrastructure is funded by ENEA, the Italian National Agency for New
Technologies, Energy and Sustainable Economic Development and by Italian and
European research programs, see http://www.cresco.enea.it/english for
information. G.~Muna\`o and G.~Milano acknowledge financial support from the Project PRIN-MIUR
2015-2016 and from the European Project Intelligent bulk MAterials for Smart TRanspOrt industries (MASTRO) 2018.


\begin{mcitethebibliography}{80}
\providecommand*{\natexlab}[1]{#1}
\providecommand*{\mciteSetBstSublistMode}[1]{}
\providecommand*{\mciteSetBstMaxWidthForm}[2]{}
\providecommand*{\mciteBstWouldAddEndPuncttrue}
  {\def\EndOfBibitem{\unskip.}}
\providecommand*{\mciteBstWouldAddEndPunctfalse}
  {\let\EndOfBibitem\relax}
\providecommand*{\mciteSetBstMidEndSepPunct}[3]{}
\providecommand*{\mciteSetBstSublistLabelBeginEnd}[3]{}
\providecommand*{\EndOfBibitem}{}
\mciteSetBstSublistMode{f}
\mciteSetBstMaxWidthForm{subitem}
{(\emph{\alph{mcitesubitemcount}})}
\mciteSetBstSublistLabelBeginEnd{\mcitemaxwidthsubitemform\space}
{\relax}{\relax}

\bibitem[Glotzer \emph{et~al.}(2017)Glotzer, Nordlander, and
  Fernandez]{Glotzer:17}
S.~C. Glotzer, P.~Nordlander and L.~E. Fernandez, \emph{ACS Nano}, 2017,
  \textbf{11}, 6505\relax
\mciteBstWouldAddEndPuncttrue
\mciteSetBstMidEndSepPunct{\mcitedefaultmidpunct}
{\mcitedefaultendpunct}{\mcitedefaultseppunct}\relax
\EndOfBibitem
\bibitem[Kumar \emph{et~al.}(2017)Kumar, Ganesan, and Riggleman]{Kumar:17}
S.~K. Kumar, V.~Ganesan and R.~A. Riggleman, \emph{J. Chem. Phys.}, 2017,
  \textbf{147}, 020901\relax
\mciteBstWouldAddEndPuncttrue
\mciteSetBstMidEndSepPunct{\mcitedefaultmidpunct}
{\mcitedefaultendpunct}{\mcitedefaultseppunct}\relax
\EndOfBibitem
\bibitem[Koo(2016)]{Koo:06}
J.~H. Koo, \emph{Polymer Nanocomposites: Processing, Characterization, and
  Application}, McGraw-Hill, 2016\relax
\mciteBstWouldAddEndPuncttrue
\mciteSetBstMidEndSepPunct{\mcitedefaultmidpunct}
{\mcitedefaultendpunct}{\mcitedefaultseppunct}\relax
\EndOfBibitem
\bibitem[Balazs \emph{et~al.}(2006)Balazs, Emrick, and Russell]{Balazs:06}
A.~C. Balazs, T.~Emrick and T.~P. Russell, \emph{Science}, 2006, \textbf{314},
  1107\relax
\mciteBstWouldAddEndPuncttrue
\mciteSetBstMidEndSepPunct{\mcitedefaultmidpunct}
{\mcitedefaultendpunct}{\mcitedefaultseppunct}\relax
\EndOfBibitem
\bibitem[Brown \emph{et~al.}(2008)Brown, Marcadon, M\`el\`e, and
  Alb\`erola]{Brown:08}
D.~Brown, V.~Marcadon, P.~M\`el\`e and N.~D. Alb\`erola, \emph{Macromolecules},
  2008, \textbf{41}, 1499\relax
\mciteBstWouldAddEndPuncttrue
\mciteSetBstMidEndSepPunct{\mcitedefaultmidpunct}
{\mcitedefaultendpunct}{\mcitedefaultseppunct}\relax
\EndOfBibitem
\bibitem[Hooper \emph{et~al.}(2004)Hooper, Schweizer, Desai, Koshy, and
  Keblinski]{Hooper:04}
J.~B. Hooper, K.~S. Schweizer, T.~G. Desai, R.~Koshy and P.~Keblinski, \emph{J.
  Chem. Phys.}, 2004, \textbf{121}, 6986\relax
\mciteBstWouldAddEndPuncttrue
\mciteSetBstMidEndSepPunct{\mcitedefaultmidpunct}
{\mcitedefaultendpunct}{\mcitedefaultseppunct}\relax
\EndOfBibitem
\bibitem[Ndoro \emph{et~al.}(2011)Ndoro, Voyiatzis, Ghanbari, Theodorou,
  B{\"o}hm, and M{\"u}ller-Plathe]{Ndoro:11}
T.~V.~M. Ndoro, E.~Voyiatzis, A.~Ghanbari, D.~N. Theodorou, M.~C. B{\"o}hm and
  F.~M{\"u}ller-Plathe, \emph{Macromolecules}, 2011, \textbf{44}, 2316\relax
\mciteBstWouldAddEndPuncttrue
\mciteSetBstMidEndSepPunct{\mcitedefaultmidpunct}
{\mcitedefaultendpunct}{\mcitedefaultseppunct}\relax
\EndOfBibitem
\bibitem[Liu \emph{et~al.}(2011)Liu, Wu, Shen, Gao, Zhang, and Cao]{Cao:11}
J.~Liu, Y.~Wu, J.~Shen, Y.~Gao, L.~Zhang and D.~Cao, \emph{Phys. Chem. Chem.
  Phys.}, 2011, \textbf{13}, 13058\relax
\mciteBstWouldAddEndPuncttrue
\mciteSetBstMidEndSepPunct{\mcitedefaultmidpunct}
{\mcitedefaultendpunct}{\mcitedefaultseppunct}\relax
\EndOfBibitem
\bibitem[Meng \emph{et~al.}(2013)Meng, Kumar, Cheng, and Grest]{Meng:13}
D.~Meng, S.~K. Kumar, S.~Cheng and G.~S. Grest, \emph{Soft Matter}, 2013,
  \textbf{9}, 5417\relax
\mciteBstWouldAddEndPuncttrue
\mciteSetBstMidEndSepPunct{\mcitedefaultmidpunct}
{\mcitedefaultendpunct}{\mcitedefaultseppunct}\relax
\EndOfBibitem
\bibitem[Martin \emph{et~al.}(2015)Martin, Mongcopa, Ashkar, Butler,
  Krishnamoorti, and Jayaraman]{Martin:15}
T.~B. Martin, K.~I.~S. Mongcopa, R.~Ashkar, P.~Butler, R.~Krishnamoorti and
  A.~Jayaraman, \emph{J. Am. Chem. Soc.}, 2015, \textbf{137}, 10624\relax
\mciteBstWouldAddEndPuncttrue
\mciteSetBstMidEndSepPunct{\mcitedefaultmidpunct}
{\mcitedefaultendpunct}{\mcitedefaultseppunct}\relax
\EndOfBibitem
\bibitem[Dukes \emph{et~al.}(2010)Dukes, Li, Lewis, Benicewicz, Schadler, and
  Kumar]{Dukes:10}
D.~Dukes, Y.~Li, S.~Lewis, B.~Benicewicz, L.~Schadler and S.~K. Kumar,
  \emph{Macromolecules}, 2010, \textbf{43}, 1564\relax
\mciteBstWouldAddEndPuncttrue
\mciteSetBstMidEndSepPunct{\mcitedefaultmidpunct}
{\mcitedefaultendpunct}{\mcitedefaultseppunct}\relax
\EndOfBibitem
\bibitem[Xie \emph{et~al.}(2013)Xie, Huang, Huang, Yang, and Jiang]{Xie:13}
L.~Y. Xie, X.~Y. Huang, Y.~H. Huang, K.~Yang and P.~K. Jiang, \emph{J. Phys.
  Chem. C}, 2013, \textbf{117}, 22525\relax
\mciteBstWouldAddEndPuncttrue
\mciteSetBstMidEndSepPunct{\mcitedefaultmidpunct}
{\mcitedefaultendpunct}{\mcitedefaultseppunct}\relax
\EndOfBibitem
\bibitem[Tang \emph{et~al.}(2014)Tang, Long, Hu, Wong, Lau, Fan, Mei, Xu, Wang,
  and Hui]{Tang:14}
C.~Y. Tang, G.~C. Long, X.~Hu, K.~W. Wong, W.~M. Lau, M.~K. Fan, J.~Mei, T.~Xu,
  B.~Wang and D.~Hui, \emph{Nanoscale}, 2014, \textbf{6}, 7877\relax
\mciteBstWouldAddEndPuncttrue
\mciteSetBstMidEndSepPunct{\mcitedefaultmidpunct}
{\mcitedefaultendpunct}{\mcitedefaultseppunct}\relax
\EndOfBibitem
\bibitem[Patel and Egorov(2004)]{Patel:04}
N.~Patel and S.~A. Egorov, \emph{J. Chem. Phys.}, 2004, \textbf{121},
  4987\relax
\mciteBstWouldAddEndPuncttrue
\mciteSetBstMidEndSepPunct{\mcitedefaultmidpunct}
{\mcitedefaultendpunct}{\mcitedefaultseppunct}\relax
\EndOfBibitem
\bibitem[Schweizer and Curro(1987)]{curro1}
K.~S. Schweizer and J.~G. Curro, \emph{Phys. Rev. Lett.}, 1987, \textbf{58},
  246\relax
\mciteBstWouldAddEndPuncttrue
\mciteSetBstMidEndSepPunct{\mcitedefaultmidpunct}
{\mcitedefaultendpunct}{\mcitedefaultseppunct}\relax
\EndOfBibitem
\bibitem[Curro and Schweizer(1987)]{curro2}
J.~G. Curro and K.~S. Schweizer, \emph{Macromolecules}, 1987, \textbf{20},
  1928\relax
\mciteBstWouldAddEndPuncttrue
\mciteSetBstMidEndSepPunct{\mcitedefaultmidpunct}
{\mcitedefaultendpunct}{\mcitedefaultseppunct}\relax
\EndOfBibitem
\bibitem[Striolo and Egorov(2007)]{Egorov:07}
A.~Striolo and S.~A. Egorov, \emph{J. Chem. Phys.}, 2007, \textbf{126},
  014902\relax
\mciteBstWouldAddEndPuncttrue
\mciteSetBstMidEndSepPunct{\mcitedefaultmidpunct}
{\mcitedefaultendpunct}{\mcitedefaultseppunct}\relax
\EndOfBibitem
\bibitem[Egorov(2008)]{Egorov:08}
S.~A. Egorov, \emph{J. Chem. Phys.}, 2008, \textbf{129}, 064901\relax
\mciteBstWouldAddEndPuncttrue
\mciteSetBstMidEndSepPunct{\mcitedefaultmidpunct}
{\mcitedefaultendpunct}{\mcitedefaultseppunct}\relax
\EndOfBibitem
\bibitem[Allegra \emph{et~al.}(2008)Allegra, Raos, and Vacatello]{Raos:08}
G.~Allegra, G.~Raos and M.~Vacatello, \emph{Prog. Polym. Sci.}, 2008,
  \textbf{33}, 683\relax
\mciteBstWouldAddEndPuncttrue
\mciteSetBstMidEndSepPunct{\mcitedefaultmidpunct}
{\mcitedefaultendpunct}{\mcitedefaultseppunct}\relax
\EndOfBibitem
\bibitem[Jancar \emph{et~al.}(2010)Jancar, Douglas, Starr, Kumar, Cassagnau,
  Lesser, Sternstein, and Buehler]{Jancar:10}
J.~Jancar, J.~F. Douglas, F.~W. Starr, S.~K. Kumar, P.~Cassagnau, A.~J. Lesser,
  S.~S. Sternstein and M.~J. Buehler, \emph{Polymer}, 2010, \textbf{51},
  3321\relax
\mciteBstWouldAddEndPuncttrue
\mciteSetBstMidEndSepPunct{\mcitedefaultmidpunct}
{\mcitedefaultendpunct}{\mcitedefaultseppunct}\relax
\EndOfBibitem
\bibitem[Ganesan and Jayaraman(2014)]{Ganesan:14}
V.~Ganesan and A.~Jayaraman, \emph{Soft Matter}, 2014, \textbf{10}, 13\relax
\mciteBstWouldAddEndPuncttrue
\mciteSetBstMidEndSepPunct{\mcitedefaultmidpunct}
{\mcitedefaultendpunct}{\mcitedefaultseppunct}\relax
\EndOfBibitem
\bibitem[Barbier \emph{et~al.}(2004)Barbier, Brown, Grillet, and
  Neyertz]{Barbier:04}
D.~Barbier, D.~Brown, A.~C. Grillet and S.~Neyertz, \emph{Macromolecules},
  2004, \textbf{37}, 4695\relax
\mciteBstWouldAddEndPuncttrue
\mciteSetBstMidEndSepPunct{\mcitedefaultmidpunct}
{\mcitedefaultendpunct}{\mcitedefaultseppunct}\relax
\EndOfBibitem
\bibitem[Eslami \emph{et~al.}(2013)Eslami, Rahimi, and
  M{\"u}ller-Plathe]{Eslami:13}
H.~Eslami, M.~Rahimi and F.~M{\"u}ller-Plathe, \emph{Macromolecules}, 2013,
  \textbf{46}, 8680\relax
\mciteBstWouldAddEndPuncttrue
\mciteSetBstMidEndSepPunct{\mcitedefaultmidpunct}
{\mcitedefaultendpunct}{\mcitedefaultseppunct}\relax
\EndOfBibitem
\bibitem[{De Nicola} \emph{et~al.}(2015){De Nicola}, Avolio, {Della Monica},
  Gentile, Cocca, Capacchione, Errico, and Milano]{Denicola:15}
A.~{De Nicola}, R.~Avolio, F.~{Della Monica}, G.~Gentile, M.~Cocca,
  C.~Capacchione, M.~E. Errico and G.~Milano, \emph{RSC Advances}, 2015,
  \textbf{5}, 71336\relax
\mciteBstWouldAddEndPuncttrue
\mciteSetBstMidEndSepPunct{\mcitedefaultmidpunct}
{\mcitedefaultendpunct}{\mcitedefaultseppunct}\relax
\EndOfBibitem
\bibitem[Karatrantos \emph{et~al.}(2015)Karatrantos, Clarke, Composto, and
  Winey]{Karatrantos:15}
A.~Karatrantos, N.~Clarke, R.~J. Composto and K.~I. Winey, \emph{Soft Matter},
  2015, \textbf{11}, 382\relax
\mciteBstWouldAddEndPuncttrue
\mciteSetBstMidEndSepPunct{\mcitedefaultmidpunct}
{\mcitedefaultendpunct}{\mcitedefaultseppunct}\relax
\EndOfBibitem
\bibitem[Reith \emph{et~al.}(2003)Reith, P{\"u}tz, and
  M{\"u}ller-Plathe]{Reith:03}
D.~Reith, M.~P{\"u}tz and F.~M{\"u}ller-Plathe, \emph{J. Comput. Chem.}, 2003,
  \textbf{24}, 1624\relax
\mciteBstWouldAddEndPuncttrue
\mciteSetBstMidEndSepPunct{\mcitedefaultmidpunct}
{\mcitedefaultendpunct}{\mcitedefaultseppunct}\relax
\EndOfBibitem
\bibitem[Huang \emph{et~al.}(2010)Huang, Faller, Do, and Moul{\`e}]{Huang:10}
D.~M. Huang, R.~Faller, K.~Do and A.~J. Moul{\`e}, \emph{J. Chem. Theory
  Comput.}, 2010, \textbf{6}, 526\relax
\mciteBstWouldAddEndPuncttrue
\mciteSetBstMidEndSepPunct{\mcitedefaultmidpunct}
{\mcitedefaultendpunct}{\mcitedefaultseppunct}\relax
\EndOfBibitem
\bibitem[Ghanbari \emph{et~al.}(2012)Ghanbari, Ndoro, Leroy, Rahimi, B{\"o}hm,
  and M{\"u}ller-Plathe]{Muller-Plathe:12}
A.~Ghanbari, T.~V.~M. Ndoro, F.~Leroy, M.~Rahimi, M.~C. B{\"o}hm and
  F.~M{\"u}ller-Plathe, \emph{Macromolecules}, 2012, \textbf{45}, 572\relax
\mciteBstWouldAddEndPuncttrue
\mciteSetBstMidEndSepPunct{\mcitedefaultmidpunct}
{\mcitedefaultendpunct}{\mcitedefaultseppunct}\relax
\EndOfBibitem
\bibitem[Shen \emph{et~al.}(2015)Shen, Liu, Li, Gao, Li, Wu, and
  Zhang]{Shen:15}
J.~Shen, J.~Liu, H.~Li, Y.~Gao, X.~Li, Y.~Wu and L.~Zhang, \emph{Phys. Chem.
  Chem. Phys.}, 2015, \textbf{17}, 7196\relax
\mciteBstWouldAddEndPuncttrue
\mciteSetBstMidEndSepPunct{\mcitedefaultmidpunct}
{\mcitedefaultendpunct}{\mcitedefaultseppunct}\relax
\EndOfBibitem
\bibitem[Shi \emph{et~al.}(2017)Shi, Qian, and Lu]{Shi:17}
R.~Shi, H.-J. Qian and Z.-Y. Lu, \emph{Phys. Chem. Chem. Phys.}, 2017,
  \textbf{19}, 16524\relax
\mciteBstWouldAddEndPuncttrue
\mciteSetBstMidEndSepPunct{\mcitedefaultmidpunct}
{\mcitedefaultendpunct}{\mcitedefaultseppunct}\relax
\EndOfBibitem
\bibitem[Kawakatsu(2004)]{Kawakatsu:04}
T.~Kawakatsu, \emph{Statistical Physics of Polymers}, Springer, Berlin,
  2004\relax
\mciteBstWouldAddEndPuncttrue
\mciteSetBstMidEndSepPunct{\mcitedefaultmidpunct}
{\mcitedefaultendpunct}{\mcitedefaultseppunct}\relax
\EndOfBibitem
\bibitem[Milano and Kawakatsu(2009)]{Milano:09}
G.~Milano and T.~Kawakatsu, \emph{J. Chem. Phys.}, 2009, \textbf{130},
  214106\relax
\mciteBstWouldAddEndPuncttrue
\mciteSetBstMidEndSepPunct{\mcitedefaultmidpunct}
{\mcitedefaultendpunct}{\mcitedefaultseppunct}\relax
\EndOfBibitem
\bibitem[Milano and Kawakatsu(2010)]{Milano:10}
G.~Milano and T.~Kawakatsu, \emph{J. Chem. Phys.}, 2010, \textbf{133},
  214102\relax
\mciteBstWouldAddEndPuncttrue
\mciteSetBstMidEndSepPunct{\mcitedefaultmidpunct}
{\mcitedefaultendpunct}{\mcitedefaultseppunct}\relax
\EndOfBibitem
\bibitem[Zhu \emph{et~al.}(2016)Zhu, Lu, Milano, Chang, and Sun]{Zhu:16}
Y.-L. Zhu, Z.-Y. Lu, G.~Milano, A.-C. Chang and Z.-Y. Sun, \emph{Phys. Chem.
  Chem. Phys.}, 2016, \textbf{18}, 9799\relax
\mciteBstWouldAddEndPuncttrue
\mciteSetBstMidEndSepPunct{\mcitedefaultmidpunct}
{\mcitedefaultendpunct}{\mcitedefaultseppunct}\relax
\EndOfBibitem
\bibitem[Zhao \emph{et~al.}(2016)Zhao, Byshkin, Cong, Kawakatsu, Guadagno, {De
  Nicola}, Yu, Milano, and Dong]{Zhao:16}
Y.~Zhao, M.~Byshkin, Y.~Cong, T.~Kawakatsu, L.~Guadagno, A.~{De Nicola}, N.~S.
  Yu, G.~Milano and B.~Dong, \emph{Nanoscale}, 2016, \textbf{8}, 15538\relax
\mciteBstWouldAddEndPuncttrue
\mciteSetBstMidEndSepPunct{\mcitedefaultmidpunct}
{\mcitedefaultendpunct}{\mcitedefaultseppunct}\relax
\EndOfBibitem
\bibitem[{De Nicola} \emph{et~al.}(2014){De Nicola}, Kawakatsu, and
  Milano]{Denicola:14}
A.~{De Nicola}, T.~Kawakatsu and G.~Milano, \emph{J. Chem. Theory Comput.},
  2014, \textbf{10}, 5651\relax
\mciteBstWouldAddEndPuncttrue
\mciteSetBstMidEndSepPunct{\mcitedefaultmidpunct}
{\mcitedefaultendpunct}{\mcitedefaultseppunct}\relax
\EndOfBibitem
\bibitem[{De Nicola} \emph{et~al.}(2016){De Nicola}, {Kawakatsu},
  {M{\"u}ller-Plathe}, and {Milano}]{Denicola:16}
A.~{De Nicola}, T.~{Kawakatsu}, F.~{M{\"u}ller-Plathe} and G.~{Milano},
  \emph{Eur. Phys. J. Special Topics}, 2016, \textbf{225}, 1817\relax
\mciteBstWouldAddEndPuncttrue
\mciteSetBstMidEndSepPunct{\mcitedefaultmidpunct}
{\mcitedefaultendpunct}{\mcitedefaultseppunct}\relax
\EndOfBibitem
\bibitem[Soares \emph{et~al.}(2017)Soares, Vanni, Milano, and
  Cascella]{Milano:17}
T.~A. Soares, S.~Vanni, G.~Milano and M.~Cascella, \emph{J. Phys. Chem. Lett.},
  2017, \textbf{8}, 3586\relax
\mciteBstWouldAddEndPuncttrue
\mciteSetBstMidEndSepPunct{\mcitedefaultmidpunct}
{\mcitedefaultendpunct}{\mcitedefaultseppunct}\relax
\EndOfBibitem
\bibitem[Lan \emph{et~al.}(2007)Lan, Francis, and Bates]{Lan:07}
Q.~Lan, L.~F. Francis and F.~S. Bates, \emph{J. Polym. Sci., Polym. Phys.},
  2007, \textbf{45}, 2284\relax
\mciteBstWouldAddEndPuncttrue
\mciteSetBstMidEndSepPunct{\mcitedefaultmidpunct}
{\mcitedefaultendpunct}{\mcitedefaultseppunct}\relax
\EndOfBibitem
\bibitem[Akcora \emph{et~al.}(2009)Akcora, Liu, Kumar, Moll, Li, Benicewicz,
  Schadler, Acechin, Panagiotopoulos, Pyramitsyn, Ganesan, Ilavsky,
  Thiyagarajan, Colby, and Douglas]{Akcora:09}
P.~Akcora, H.~Liu, S.~K. Kumar, J.~Moll, Y.~Li, B.~C. Benicewicz, L.~S.
  Schadler, D.~Acechin, A.~Z. Panagiotopoulos, V.~Pyramitsyn, V.~Ganesan,
  J.~Ilavsky, P.~Thiyagarajan, R.~H. Colby and J.~F. Douglas, \emph{Nat.
  Mater.}, 2009, \textbf{8}, 354\relax
\mciteBstWouldAddEndPuncttrue
\mciteSetBstMidEndSepPunct{\mcitedefaultmidpunct}
{\mcitedefaultendpunct}{\mcitedefaultseppunct}\relax
\EndOfBibitem
\bibitem[Chevigny \emph{et~al.}(2010)Chevigny, Jestin, Gigmes, Schweins, {Di
  Cola}, Dalmas, Bertin, and Bou{\`e}]{Chevigny:10}
C.~Chevigny, J.~Jestin, D.~Gigmes, R.~Schweins, E.~{Di Cola}, F.~Dalmas,
  D.~Bertin and F.~Bou{\`e}, \emph{Macromolecules}, 2010, \textbf{43},
  4833\relax
\mciteBstWouldAddEndPuncttrue
\mciteSetBstMidEndSepPunct{\mcitedefaultmidpunct}
{\mcitedefaultendpunct}{\mcitedefaultseppunct}\relax
\EndOfBibitem
\bibitem[Chevigny \emph{et~al.}(2011)Chevigny, Dalmas, {Di Cola}, Gigmes,
  Bertin, Boue, and Jestin]{Chevigny:11}
C.~Chevigny, F.~Dalmas, E.~{Di Cola}, D.~Gigmes, D.~Bertin, F.~Boue and
  J.~Jestin, \emph{Macromolecules}, 2011, \textbf{44}, 122\relax
\mciteBstWouldAddEndPuncttrue
\mciteSetBstMidEndSepPunct{\mcitedefaultmidpunct}
{\mcitedefaultendpunct}{\mcitedefaultseppunct}\relax
\EndOfBibitem
\bibitem[Sunday \emph{et~al.}(2012)Sunday, Ilavsky, and Green]{Sunday:12}
D.~Sunday, J.~Ilavsky and D.~L. Green, \emph{Macromolecules}, 2012,
  \textbf{45}, 4007\relax
\mciteBstWouldAddEndPuncttrue
\mciteSetBstMidEndSepPunct{\mcitedefaultmidpunct}
{\mcitedefaultendpunct}{\mcitedefaultseppunct}\relax
\EndOfBibitem
\bibitem[Ghanbari \emph{et~al.}(2013)Ghanbari, Rahimi, and
  Dehghany]{Ghanbari:13}
A.~Ghanbari, M.~Rahimi and J.~Dehghany, \emph{J. Phys. Chem. C}, 2013,
  \textbf{117}, 25069\relax
\mciteBstWouldAddEndPuncttrue
\mciteSetBstMidEndSepPunct{\mcitedefaultmidpunct}
{\mcitedefaultendpunct}{\mcitedefaultseppunct}\relax
\EndOfBibitem
\bibitem[Voyiatzis \emph{et~al.}(2016)Voyiatzis, M{\"u}ller-Plathe, and
  B{\"o}hm]{Voyiatzis:16}
E.~Voyiatzis, F.~M{\"u}ller-Plathe and M.~C. B{\"o}hm, \emph{Polymer}, 2016,
  \textbf{101}, 107\relax
\mciteBstWouldAddEndPuncttrue
\mciteSetBstMidEndSepPunct{\mcitedefaultmidpunct}
{\mcitedefaultendpunct}{\mcitedefaultseppunct}\relax
\EndOfBibitem
\bibitem[Pfaller \emph{et~al.}(2016)Pfaller, Possart, Steinmann, Rahimi,
  M{\"u}ller-Plathe, and B{\"o}hm]{Pfaller:16}
S.~Pfaller, G.~Possart, P.~Steinmann, M.~Rahimi, F.~M{\"u}ller-Plathe and M.~C.
  B{\"o}hm, \emph{Phys. Rev. E}, 2016, \textbf{93}, 052505\relax
\mciteBstWouldAddEndPuncttrue
\mciteSetBstMidEndSepPunct{\mcitedefaultmidpunct}
{\mcitedefaultendpunct}{\mcitedefaultseppunct}\relax
\EndOfBibitem
\bibitem[Rahimi \emph{et~al.}(2012)Rahimi, Iriarte-Carretero, Ghanbari,
  B{\"o}hm, and M{\"u}ller-Plathe]{Rahimi:12}
M.~Rahimi, I.~Iriarte-Carretero, A.~Ghanbari, M.~C. B{\"o}hm and
  F.~M{\"u}ller-Plathe, \emph{Nanotechnology}, 2012, \textbf{23}, 305702\relax
\mciteBstWouldAddEndPuncttrue
\mciteSetBstMidEndSepPunct{\mcitedefaultmidpunct}
{\mcitedefaultendpunct}{\mcitedefaultseppunct}\relax
\EndOfBibitem
\bibitem[Ndoro \emph{et~al.}(2012)Ndoro, B{\"o}hm, and
  M{\"u}ller-Plathe]{Ndoro:12}
T.~V.~M. Ndoro, M.~C. B{\"o}hm and F.~M{\"u}ller-Plathe, \emph{Macromolecules},
  2012, \textbf{45}, 171\relax
\mciteBstWouldAddEndPuncttrue
\mciteSetBstMidEndSepPunct{\mcitedefaultmidpunct}
{\mcitedefaultendpunct}{\mcitedefaultseppunct}\relax
\EndOfBibitem
\bibitem[Frischknecht and Yethiraj(2011)]{Yeth:11}
A.~L. Frischknecht and A.~Yethiraj, \emph{J. Chem. Phys.}, 2011, \textbf{134},
  174901\relax
\mciteBstWouldAddEndPuncttrue
\mciteSetBstMidEndSepPunct{\mcitedefaultmidpunct}
{\mcitedefaultendpunct}{\mcitedefaultseppunct}\relax
\EndOfBibitem
\bibitem[Martin \emph{et~al.}(2013)Martin, Dodd, and Jayaraman]{Martin:13}
T.~B. Martin, P.~M. Dodd and A.~Jayaraman, \emph{Phys. Rev. Lett.}, 2013,
  \textbf{110}, 018301\relax
\mciteBstWouldAddEndPuncttrue
\mciteSetBstMidEndSepPunct{\mcitedefaultmidpunct}
{\mcitedefaultendpunct}{\mcitedefaultseppunct}\relax
\EndOfBibitem
\bibitem[Cerd{\`a} \emph{et~al.}(2003)Cerd{\`a}, Sintes, and Toral]{Cerda:03}
J.~J. Cerd{\`a}, T.~Sintes and R.~Toral, \emph{Macromolecules}, 2003,
  \textbf{36}, 1407\relax
\mciteBstWouldAddEndPuncttrue
\mciteSetBstMidEndSepPunct{\mcitedefaultmidpunct}
{\mcitedefaultendpunct}{\mcitedefaultseppunct}\relax
\EndOfBibitem
\bibitem[Smith \emph{et~al.}(2003)Smith, Bedrov, and Smith]{Smith:03}
J.~S. Smith, D.~Bedrov and G.~D. Smith, \emph{Compos. Sci. Technol.}, 2003,
  \textbf{63}, 1599\relax
\mciteBstWouldAddEndPuncttrue
\mciteSetBstMidEndSepPunct{\mcitedefaultmidpunct}
{\mcitedefaultendpunct}{\mcitedefaultseppunct}\relax
\EndOfBibitem
\bibitem[Marla and Meredith(2006)]{Marla:06}
K.~T. Marla and J.~C. Meredith, \emph{J. Chem. Theory Comput.}, 2006,
  \textbf{2}, 1624\relax
\mciteBstWouldAddEndPuncttrue
\mciteSetBstMidEndSepPunct{\mcitedefaultmidpunct}
{\mcitedefaultendpunct}{\mcitedefaultseppunct}\relax
\EndOfBibitem
\bibitem[Smith and Bedrov(2009)]{Smith:09}
G.~D. Smith and D.~Bedrov, \emph{Langmuir}, 2009, \textbf{25}, 11239\relax
\mciteBstWouldAddEndPuncttrue
\mciteSetBstMidEndSepPunct{\mcitedefaultmidpunct}
{\mcitedefaultendpunct}{\mcitedefaultseppunct}\relax
\EndOfBibitem
\bibitem[{Lo Verso} \emph{et~al.}(2011){Lo Verso}, Yelash, Egorov, and
  Binder]{Loverso:11}
F.~{Lo Verso}, L.~Yelash, S.~A. Egorov and K.~Binder, \emph{J. Chem. Phys.},
  2011, \textbf{135}, 214902\relax
\mciteBstWouldAddEndPuncttrue
\mciteSetBstMidEndSepPunct{\mcitedefaultmidpunct}
{\mcitedefaultendpunct}{\mcitedefaultseppunct}\relax
\EndOfBibitem
\bibitem[Meng \emph{et~al.}(2012)Meng, Kumar, Lane, and Grest]{Meng:12}
D.~Meng, S.~K. Kumar, J.~M.~D. Lane and G.~S. Grest, \emph{Soft Matter}, 2012,
  \textbf{8}, 5002\relax
\mciteBstWouldAddEndPuncttrue
\mciteSetBstMidEndSepPunct{\mcitedefaultmidpunct}
{\mcitedefaultendpunct}{\mcitedefaultseppunct}\relax
\EndOfBibitem
\bibitem[Baran and Sokolowski(2017)]{Baran:17}
L.~Baran and S.~Sokolowski, \emph{J. Chem. Phys.}, 2017, \textbf{147},
  044903\relax
\mciteBstWouldAddEndPuncttrue
\mciteSetBstMidEndSepPunct{\mcitedefaultmidpunct}
{\mcitedefaultendpunct}{\mcitedefaultseppunct}\relax
\EndOfBibitem
\bibitem[Rank and Baker(1997)]{Rank:97}
J.~A. Rank and D.~Baker, \emph{Protein Sci.}, 1997, \textbf{6}, 347\relax
\mciteBstWouldAddEndPuncttrue
\mciteSetBstMidEndSepPunct{\mcitedefaultmidpunct}
{\mcitedefaultendpunct}{\mcitedefaultseppunct}\relax
\EndOfBibitem
\bibitem[Czaplewski \emph{et~al.}(2000)Czaplewski, Rodziewicz-Motowidlo, Liwo,
  Ripoll, Wawak, and Scheraga]{Czaplewski:00}
C.~Czaplewski, S.~Rodziewicz-Motowidlo, A.~Liwo, D.~R. Ripoll, R.~J. Wawak and
  H.~A. Scheraga, \emph{Protein Sci.}, 2000, \textbf{9}, 1235\relax
\mciteBstWouldAddEndPuncttrue
\mciteSetBstMidEndSepPunct{\mcitedefaultmidpunct}
{\mcitedefaultendpunct}{\mcitedefaultseppunct}\relax
\EndOfBibitem
\bibitem[Czaplewski \emph{et~al.}(2003)Czaplewski, Rodziewicz-Motowidlo, Dabal,
  Liwo, Ripoll, and Scheraga]{Czaplewski:03}
C.~Czaplewski, S.~Rodziewicz-Motowidlo, M.~Dabal, A.~Liwo, D.~R. Ripoll and
  H.~A. Scheraga, \emph{Biophys. Chem.}, 2003, \textbf{105}, 339\relax
\mciteBstWouldAddEndPuncttrue
\mciteSetBstMidEndSepPunct{\mcitedefaultmidpunct}
{\mcitedefaultendpunct}{\mcitedefaultseppunct}\relax
\EndOfBibitem
\bibitem[Qian \emph{et~al.}(2008)Qian, Carbone, Chen, Karimi-Varzaneh, Liew,
  and M{\"u}ller-Plathe]{Qian:08}
H.-J. Qian, P.~Carbone, X.~Chen, H.~A. Karimi-Varzaneh, C.~C. Liew and
  F.~M{\"u}ller-Plathe, \emph{Macromolecules}, 2008, \textbf{41}, 9919\relax
\mciteBstWouldAddEndPuncttrue
\mciteSetBstMidEndSepPunct{\mcitedefaultmidpunct}
{\mcitedefaultendpunct}{\mcitedefaultseppunct}\relax
\EndOfBibitem
\bibitem[Sides \emph{et~al.}(2006)Sides, Kim, Kramer, and
  Fredrickson]{Sides:06}
S.~W. Sides, B.~J. Kim, E.~J. Kramer and G.~H. Fredrickson, \emph{Phys. Rev.
  Lett.}, 2006, \textbf{96}, 250601\relax
\mciteBstWouldAddEndPuncttrue
\mciteSetBstMidEndSepPunct{\mcitedefaultmidpunct}
{\mcitedefaultendpunct}{\mcitedefaultseppunct}\relax
\EndOfBibitem
\bibitem[Zhao \emph{et~al.}(2012)Zhao, {De Nicola}, Kawakatsu, and
  Milano]{Occam}
Y.~Zhao, A.~{De Nicola}, T.~Kawakatsu and G.~Milano, \emph{J. Comput. Chem.},
  2012, \textbf{33}, 868\relax
\mciteBstWouldAddEndPuncttrue
\mciteSetBstMidEndSepPunct{\mcitedefaultmidpunct}
{\mcitedefaultendpunct}{\mcitedefaultseppunct}\relax
\EndOfBibitem
\bibitem[Martinez \emph{et~al.}(2009)Martinez, Andrade, Birgin, and
  Martinez]{Packmol}
L.~Martinez, R.~Andrade, E.~G. Birgin and J.~M. Martinez, \emph{J. Comput.
  Chem.}, 2009, \textbf{30}, 2157\relax
\mciteBstWouldAddEndPuncttrue
\mciteSetBstMidEndSepPunct{\mcitedefaultmidpunct}
{\mcitedefaultendpunct}{\mcitedefaultseppunct}\relax
\EndOfBibitem
\bibitem[Hamaker(1937)]{Hamaker:37}
H.~C. Hamaker, \emph{Physica}, 1937, \textbf{4}, 1058\relax
\mciteBstWouldAddEndPuncttrue
\mciteSetBstMidEndSepPunct{\mcitedefaultmidpunct}
{\mcitedefaultendpunct}{\mcitedefaultseppunct}\relax
\EndOfBibitem
\bibitem[Muna{\`o} \emph{et~al.}(2018)Muna{\`o}, Correa, Pizzirusso, and
  Milano]{Munao:18}
G.~Muna{\`o}, A.~Correa, A.~Pizzirusso and G.~Milano, \emph{EPJ E}, 2018,
  \textbf{41}, 38\relax
\mciteBstWouldAddEndPuncttrue
\mciteSetBstMidEndSepPunct{\mcitedefaultmidpunct}
{\mcitedefaultendpunct}{\mcitedefaultseppunct}\relax
\EndOfBibitem
\bibitem[Hooper and Schweizer(2005)]{Hooper:05}
J.~B. Hooper and K.~S. Schweizer, \emph{Macromolecules}, 2005, \textbf{38},
  8858\relax
\mciteBstWouldAddEndPuncttrue
\mciteSetBstMidEndSepPunct{\mcitedefaultmidpunct}
{\mcitedefaultendpunct}{\mcitedefaultseppunct}\relax
\EndOfBibitem
\bibitem[Bedrov \emph{et~al.}(2003)Bedrov, Smith, and Smith]{Bedrov:03}
D.~Bedrov, G.~D. Smith and J.~S. Smith, \emph{J. Chem. Phys.}, 2003,
  \textbf{119}, 10438\relax
\mciteBstWouldAddEndPuncttrue
\mciteSetBstMidEndSepPunct{\mcitedefaultmidpunct}
{\mcitedefaultendpunct}{\mcitedefaultseppunct}\relax
\EndOfBibitem
\bibitem[McQuarrie(1976)]{McQuarrie:76}
D.~A. McQuarrie, \emph{Statistical Mechanics}, Harper Collins, New York,
  1976\relax
\mciteBstWouldAddEndPuncttrue
\mciteSetBstMidEndSepPunct{\mcitedefaultmidpunct}
{\mcitedefaultendpunct}{\mcitedefaultseppunct}\relax
\EndOfBibitem
\bibitem[Kumar \emph{et~al.}(2013)Kumar, Jouault, Benicewicz, and
  Neely]{Kumar:13}
S.~K. Kumar, N.~Jouault, B.~Benicewicz and T.~Neely, \emph{Macromolecules},
  2013, \textbf{46}, 3199\relax
\mciteBstWouldAddEndPuncttrue
\mciteSetBstMidEndSepPunct{\mcitedefaultmidpunct}
{\mcitedefaultendpunct}{\mcitedefaultseppunct}\relax
\EndOfBibitem
\bibitem[Srivastava \emph{et~al.}(2012)Srivastava, Agarwal, and
  Archer]{Archer:12}
S.~Srivastava, P.~Agarwal and L.~A. Archer, \emph{Langmuir}, 2012, \textbf{28},
  6276\relax
\mciteBstWouldAddEndPuncttrue
\mciteSetBstMidEndSepPunct{\mcitedefaultmidpunct}
{\mcitedefaultendpunct}{\mcitedefaultseppunct}\relax
\EndOfBibitem
\bibitem[Hall \emph{et~al.}(2010)Hall, Jayaraman, and Schweizer]{Hall:10}
L.~M. Hall, A.~Jayaraman and K.~S. Schweizer, \emph{Curr. Opin. Solid State
  Mater. Sci.}, 2010, \textbf{14}, 38\relax
\mciteBstWouldAddEndPuncttrue
\mciteSetBstMidEndSepPunct{\mcitedefaultmidpunct}
{\mcitedefaultendpunct}{\mcitedefaultseppunct}\relax
\EndOfBibitem
\bibitem[Schmidle \emph{et~al.}(2012)Schmidle, Hall, Velev, and Klapp]{Hall:12}
H.~Schmidle, C.~K. Hall, O.~D. Velev and S.~H.~L. Klapp, \emph{Soft Matter},
  2012, \textbf{8}, 1521\relax
\mciteBstWouldAddEndPuncttrue
\mciteSetBstMidEndSepPunct{\mcitedefaultmidpunct}
{\mcitedefaultendpunct}{\mcitedefaultseppunct}\relax
\EndOfBibitem
\bibitem[Bachhar \emph{et~al.}(2017)Bachhar, Jiao, Asai, Akcora,
  Bandyopadhyaya, and Kumar]{Bachhar:17}
N.~Bachhar, Y.~Jiao, M.~Asai, P.~Akcora, R.~Bandyopadhyaya and S.~K. Kumar,
  \emph{Macromolecules}, 2017, \textbf{50}, 7730\relax
\mciteBstWouldAddEndPuncttrue
\mciteSetBstMidEndSepPunct{\mcitedefaultmidpunct}
{\mcitedefaultendpunct}{\mcitedefaultseppunct}\relax
\EndOfBibitem
\bibitem[Stradner \emph{et~al.}(2004)Stradner, Sedgwick, Cardinaux, Poon,
  Egelhaaf, and Schurtenberger]{Stradner}
A.~Stradner, H.~Sedgwick, F.~Cardinaux, W.~C. Poon, S.~U. Egelhaaf and
  P.~Schurtenberger, \emph{Nature}, 2004, \textbf{432}, 492\relax
\mciteBstWouldAddEndPuncttrue
\mciteSetBstMidEndSepPunct{\mcitedefaultmidpunct}
{\mcitedefaultendpunct}{\mcitedefaultseppunct}\relax
\EndOfBibitem
\bibitem[Preisler \emph{et~al.}(2014)Preisler, Vissers, Muna{\`o}, Smallenburg,
  and Sciortino]{Preisler:14}
Z.~Preisler, T.~Vissers, G.~Muna{\`o}, F.~Smallenburg and F.~Sciortino,
  \emph{Soft Matter}, 2014, \textbf{10}, 5121\relax
\mciteBstWouldAddEndPuncttrue
\mciteSetBstMidEndSepPunct{\mcitedefaultmidpunct}
{\mcitedefaultendpunct}{\mcitedefaultseppunct}\relax
\EndOfBibitem
\bibitem[Sciortino \emph{et~al.}(2004)Sciortino, Mossa, Zaccarelli, and
  Tartaglia]{Attr_Rep}
F.~Sciortino, S.~Mossa, E.~Zaccarelli and P.~Tartaglia, \emph{Phys. Rev.
  Lett.}, 2004, \textbf{93}, 055701\relax
\mciteBstWouldAddEndPuncttrue
\mciteSetBstMidEndSepPunct{\mcitedefaultmidpunct}
{\mcitedefaultendpunct}{\mcitedefaultseppunct}\relax
\EndOfBibitem
\bibitem[Lane and Grest(2010)]{Lane:10}
J.~M.~D. Lane and G.~S. Grest, \emph{Phys. Rev. Lett.}, 2010, \textbf{104},
  235501\relax
\mciteBstWouldAddEndPuncttrue
\mciteSetBstMidEndSepPunct{\mcitedefaultmidpunct}
{\mcitedefaultendpunct}{\mcitedefaultseppunct}\relax
\EndOfBibitem
\bibitem[Bozorgui \emph{et~al.}(2013)Bozorgui, Meng, Kumar, Chakravarty, and
  Cacciuto]{Bozorgui:13}
B.~Bozorgui, D.~Meng, S.~K. Kumar, C.~Chakravarty and A.~Cacciuto, \emph{Nano
  Lett.}, 2013, \textbf{13}, 2732\relax
\mciteBstWouldAddEndPuncttrue
\mciteSetBstMidEndSepPunct{\mcitedefaultmidpunct}
{\mcitedefaultendpunct}{\mcitedefaultseppunct}\relax
\EndOfBibitem
\bibitem[Ponti and {et al}(2014)]{Cresco}
G.~Ponti and {et al}, \emph{Proceedings of the 2014 International Conference on
  High Performance Computing and Simulation}, 2014, \textbf{6903807},
  1030\relax
\mciteBstWouldAddEndPuncttrue
\mciteSetBstMidEndSepPunct{\mcitedefaultmidpunct}
{\mcitedefaultendpunct}{\mcitedefaultseppunct}\relax
\EndOfBibitem
\end{mcitethebibliography}
%
\providecommand*{\mcitethebibliography}{\thebibliography}
\csname @ifundefined\endcsname{endmcitethebibliography}
{\let\endmcitethebibliography\endthebibliography}{}

\end{document}